# Bulk synthesis of Zn$_3$WN$_4$ via solid-state metathesis

Christopher L. Rom[1], Shaun O'Donnell[1,2], Kayla Huang[1,3], Ryan A. Klein[1,4], Morgan J. Kramer[4,5], Rebecca W. Smaha[1], Andriy Zakutayev[1,*]

[1] Materials, Chemical, and Computational Science, National Renewable Energy Laboratory, Golden, CO, 80401, USA
[2] Department of Chemistry, Colorado State University, Fort Collins, CO, 80523, USA
[3] University of Illinois Urbana-Champaign, Champaign, IL, 61801, USA
[4] Center for Neutron Research, National Institute of Standards and Technology, Gaithersburg, MD, 20899, USA
[5] Department of Chemistry, Southern Methodist University, Dallas, TX, 75275 USA

[*] Corresponding author: andriy.zakutayev@nrel.gov

## TOC figure

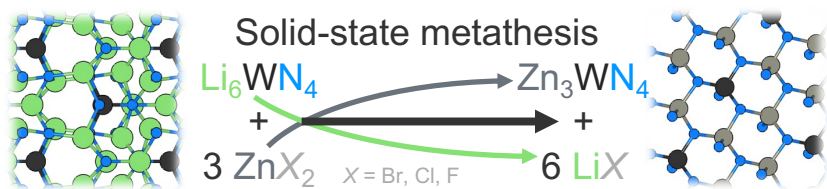

## Abstract

Ternary nitrides are of growing technological importance, with applications as semiconductors, catalysts, and magnetic materials; however, new synthetic tools are needed to advance materials discovery efforts. Here, we show that $Zn_3WN_4$ can be synthesized via metathesis reactions between $Li_6WN_4$ and $ZnX_2$ ($X$ = Br, Cl, F). *In situ* synchrotron powder X-ray diffraction and differential scanning calorimetry show that the reaction onset is correlated with the $ZnX_2$ melting point and that product purity is inversely correlated with the reaction's exothermicity. High resolution synchrotron powder X-ray diffraction measurements show that this bulk synthesis produces a structure with substantial cation ordering, as opposed to the disordered structure initially discovered via thin film sputtering. Diffuse reflectance spectroscopy reveals that $Zn_3WN_4$ powders exhibit two optical absorption onsets at ~2.5 eV and ~4.0 eV, indicating wide-bandgap semiconducting behavior and suggesting a small amount of structural disorder. We hypothesize that this synthesis strategy is generalizable because many potential Li-$M$-N precursors (where $M$ is a metal) are available for synthesizing new ternary nitride materials. This work introduces a promising synthesis strategy that will accelerate the discovery of novel functional ternary nitrides and other currently inaccessible materials.

## Introduction

Ternary nitrides are a promising class of semiconducting materials,[1] yet relatively few are known. This dearth of nitrides is primarily due to the synthetic challenges of realizing these materials from elemental metal (or binary) precursors and dinitrogen gas.[1–4] Molecular (di)nitrogen, $N_2$, is highly stable, and high temperatures are needed to break the strong N≡N triple bond (945 kJ/mol).[5] High temperatures are also needed to drive diffusion, as nitrides tend to have high cohesive energies (i.e., strong $M$–N bonds) and slow diffusion.[6–8] Moreover, entropic penalties disfavor nitride incorporation in solids (i.e., gaseous $N_2$ is favored). Finding a synthesis temperature that is hot enough for reactivity but cool enough to avoid decomposition is therefore challenging. Adding to the difficulty, $O_2$ is more reactive towards most metals than $N_2$, so syntheses must be conducted in rigorously air-free conditions to avoid the formation of oxide impurities. Consequently, the number of known ternary nitrides lags behind the ternary oxides by an order of magnitude.[1–4] Developing new synthesis methods will help narrow this disparity, and in doing so, discover new materials upon which improved technologies can be built.

Zn-containing ternary nitrides epitomize the promising applications and synthetic challenges of this class of materials. Fully nitridized compounds like $ZnSnN_2$ and $Zn_3WN_4$ (with metals in the highest oxidation state) are of interest as semiconductors for their high earth abundance and tunable bandgaps (spanning ca. 1 eV for $ZnSnN_2$ to 4 eV for $Zn_3WN_4$).[9,10] However, the bulk synthesis techniques that have been reported for Zn-$M$-N phases are limited to traditional ceramic methods (i.e., metals + $N_2$ or $NH_3$ at high temperatures) or high-pressure solid state metathesis reactions (e.g., 2 $Li_3N$ + $ZnF_2$ + $SnF_4$ → $ZnSnN_2$ + 6 LiF).[11,12] These bulk methods have only produced fully nitridized phases when $M$ is a main group element (i.e., $LiZnN$, $Ca_2ZnN_2$, $Sr_2ZnN_2$, $Ba_2ZnN_2$, $ZnSiN_2$, $ZnGeN_2$, $ZnSnN_2$).[7,9–19] When transition metals are used in bulk syntheses, they tend to form sub-nitrides: e.g., $Ti_3ZnN_{0.5}$, $V_3Zn_2N$, $Ti_2ZnN$, $Mn_3ZnN$, and $Fe_3ZnN$.[20–25] The nitrogen-poor nature of these materials stems from the challenges described above (i.e., $N_2$ stability, slow diffusion). Synthesizing fully-nitridized Zn-$M$-N (where $M$ is a transition metal) in bulk would advance technologies in which thin film nitrides have already shown promise, like photoelectrochemical energy conversion ($ZnTiN_2$),[26] transparent conducting oxides ($ZnZrN_2$),[27] and non-linear optics ($Zn_3WN_4$).[28] However, no bulk synthesis methods have been reported for fully nitridized Zn-*M*-N ternaries.

Synthesizing Zn-*M*-N ternary nitrides via traditional methods is difficult. Many transition metals are highly refractory, meaning high temperatures would likely be needed for interdiffusion of reactants. However, Zn has a low melting point (419 °C) and a relatively low boiling point (907 °C),[29] meaning that high temperatures would volatilize Zn away from the other metal unless special measures were taken (e.g., high pressure, closed vessels). Forming binary nitrides to use as precursors instead of metals is also challenging: Zn (like other late-transition metals) does not react with $N_2$ at elevated temperatures, so $Zn_3N_2$ must be synthesized under ammonia.[30] And as noted in thin film work, fully nitridized transition metal Zn-*M*-N phases have low decomposition temperatures on the order of 600-700 °C.[3,26,27,31] These challenges mean that bulk synthesis of Zn-*M*-N from the elements or binaries would likely proceed only at low temperatures and extremely slowly, unless special high-pressure methods were employed (e.g., ammonothermal synthesis,[12] diamond anvil cell synthesis,[32] etc.).

Metathesis reactions (i.e., ion exchange reactions) are one promising way to circumvent the challenge of diffusion in the solid state.[33] To synthesize nitrides, this strategy starts with one nitrogen-containing precursor and one halide precursor, rather than elements or binary nitrides. The balanced reaction targets the desired phase along with a byproduct (often a halide salt). The formation of this byproduct provides a large thermodynamic driving force for the reaction and (ideally) can be washed away post-reaction. For example, Kaner et al. showed that mixing $Li_3N$ with metal chlorides would produce LiCl in explosively exothermic metathesis reactions that yielded a range of binary nitrides[34–44] and some ternary nitrides.[45,46] Alternatively, less exothermic reactions can be conducted with greater synthetic control,[47–51] including low-temperature topotactic reactions ($T_{rxn}$ ca. 200–400 °C).[52–54] As for Zn-*M*-N compounds, $ZnSnN_2$ and $ZnSiN_2$ have been made using high pressure metathesis reactions, where the pressure is necessary to avoid gaseous $N_2$ loss.[11,19] Metathesis is well known for "turning down the heat" in solid state synthesis[55] but is underutilized for synthesizing nitrides.

Here, we synthesize $Zn_3WN_4$ via a near-topotactic metathesis reaction between $Li_6WN_4$ and $ZnX_2$ (*X* = Br, Cl, F) at 300 °C and ambient pressure. *In situ* synchrotron powder X-ray diffraction (SPXRD) paired with differential scanning calorimetry (DSC) measurements reveal the reaction pathways and show that using a $ZnBr_2$ precursor is preferable over the fluoride or chloride analogs. High resolution SPXRD measurements indicate that the $Zn_3WN_4$ product is a mostly cation-ordered structure in space group *Pmn*$2_1$. We report some preliminary properties characterizations for $Zn_3WN_4$, revealing optical absorption onsets near 2.5 eV and 4.0 eV, as well as paramagnetism consistent with some degree of disorder and off-stoichiometry. The reaction is near-topotactic, in that the structures of the $Li_6WN_4$ precursor and the $Zn_3WN_4$ product are related by a shift in anion layers but the [$WN_4$] tetrahedral unit is preserved. Using this synthesis approach, we also synthesized $Zn_3MoN_4$, albeit with lower levels of purity in our un-optimized reactions. This work demonstrates the viability of Li-*M*-N phases as metathesis precursors to synthesize other ternary nitride compounds, expanding the toolkit for materials discovery.

## Results and Discussion

*In situ* SPXRD measurements

Zn$_3$WN$_4$ was successfully synthesized via metathesis (ion exchange) reactions. The net reaction is:

Li$_6$WN$_4$ + 3 ZnX$_2$ → Zn$_3$WN$_4$ + 6 LiX (X = Br, Cl, F)

*In situ* variable temperature SPXRD measurements reveal that Li$_6$WN$_4$ directly converts to Zn$_3$WN$_4$ without intermediate crystalline phases or solid solution behavior as a function of temperature (Figure 1, Figures S2, S3). However, the halide precursor exerts an influence on the reaction kinetics and thermodynamics, which ultimately impact the reaction pathway and final product purity.

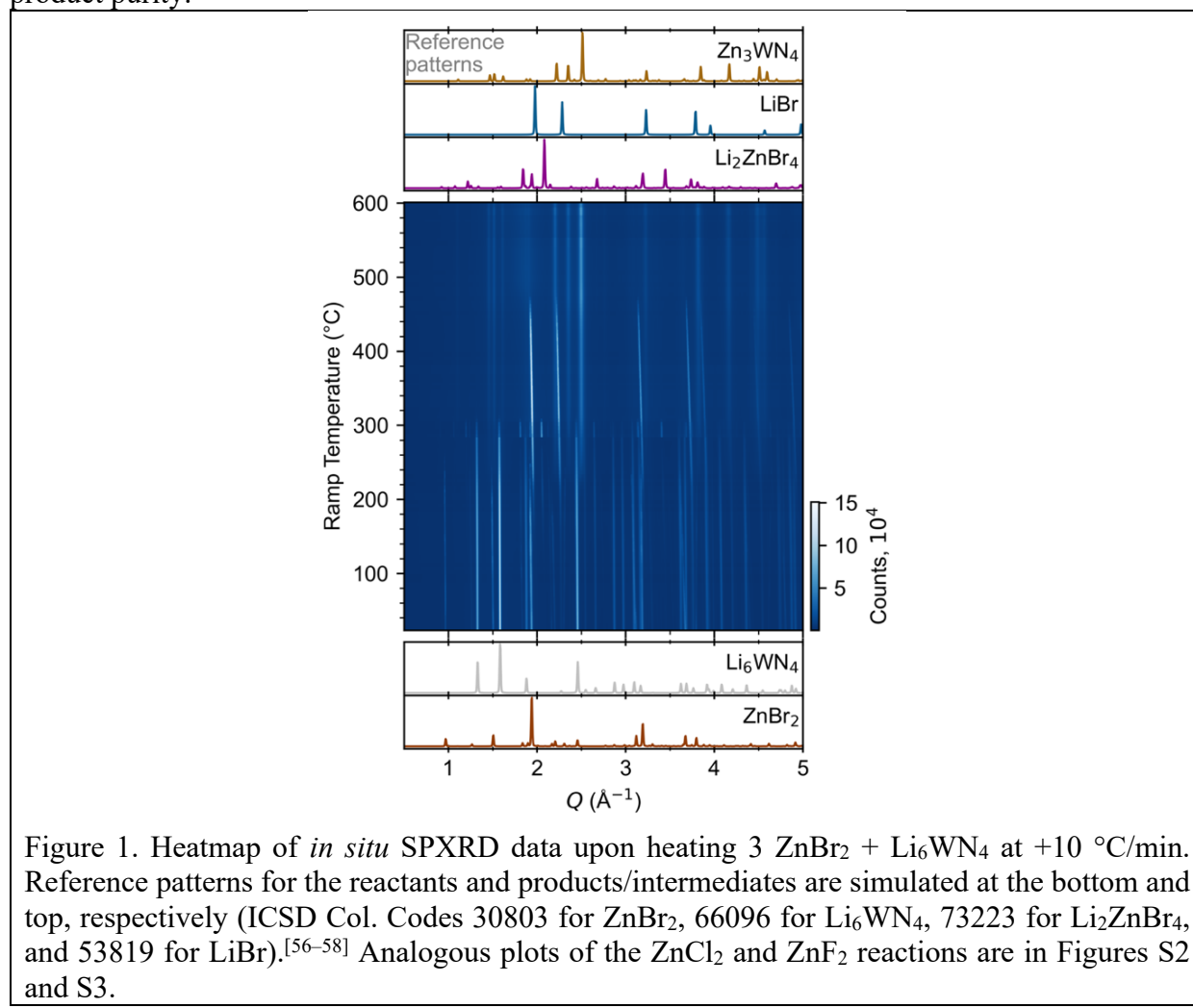

Figure 1. Heatmap of *in situ* SPXRD data upon heating 3 ZnBr$_2$ + Li$_6$WN$_4$ at +10 °C/min. Reference patterns for the reactants and products/intermediates are simulated at the bottom and top, respectively (ICSD Col. Codes 30803 for ZnBr$_2$, 66096 for Li$_6$WN$_4$, 73223 for Li$_2$ZnBr$_4$, and 53819 for LiBr).[56–58] Analogous plots of the ZnCl$_2$ and ZnF$_2$ reactions are in Figures S2 and S3.

*In situ* SPXRD measurements reveal that the reaction pathway proceeds without intermediate nitrides between Li$_6$WN$_4$ and Zn$_3$WN$_4$. Figure 1 shows a heatmap for X = Br as an example; subsequent examination revealed that it leads to the most phase pure product. The reaction of Li$_6$WN$_4$ + 3 ZnBr$_2$ → Zn$_3$WN$_4$ + 6 LiBr initiates near 170 °C and proceeds to completion within the 14 minutes of ramp time up to 310 °C. Near 170 °C, the Bragg peaks arising from crystalline Li$_6$WN$_4$ and ZnBr$_2$ begin to gradually decline in intensity. Shortly thereafter, new sets of Bragg peaks that can be indexed to LiBr, Li$_2$ZnBr$_4$, and Zn$_3$WN$_4$ begin growing in intensity in the patterns. The Bragg peaks corresponding to Li$_2$ZnBr$_4$ gradually decrease in intensity between 210 °C and 270 °C, increase dramatically in intensity at 275 °C, and then disappear entirely at 305 °C. Such fluctuations may stem from crystal nucleation and growth within the capillary, especially given the small spot size of the synchrotron X-ray beam, possibly combined with crystallite motion in a liquid-like medium. Diffraction images show spotty diffraction patterns, consistent with crystallite growth. These data indicate that the synthesis proceeds directly via Li$_6$WN$_4$ + 3 ZnBr$_2$ → Zn$_3$WN$_4$ + 6 LiBr. While this process occurs, an incidental reaction between the metal halides also occurs: 2 LiBr + ZnBr$_2$ → Li$_2$ZnBr$_4$. We do not observe signs of a crystalline theoretically-predicted LiZn$_4$W$_2$N$_7$ structure,[59] although this does not rule out the presence or synthesizability of such a phase. Similar trends are noted with the ZnCl$_2$ and ZnF$_2$ reactions (Figures S2, S3), as shown by sequential Rietveld analysis (Figure 2).

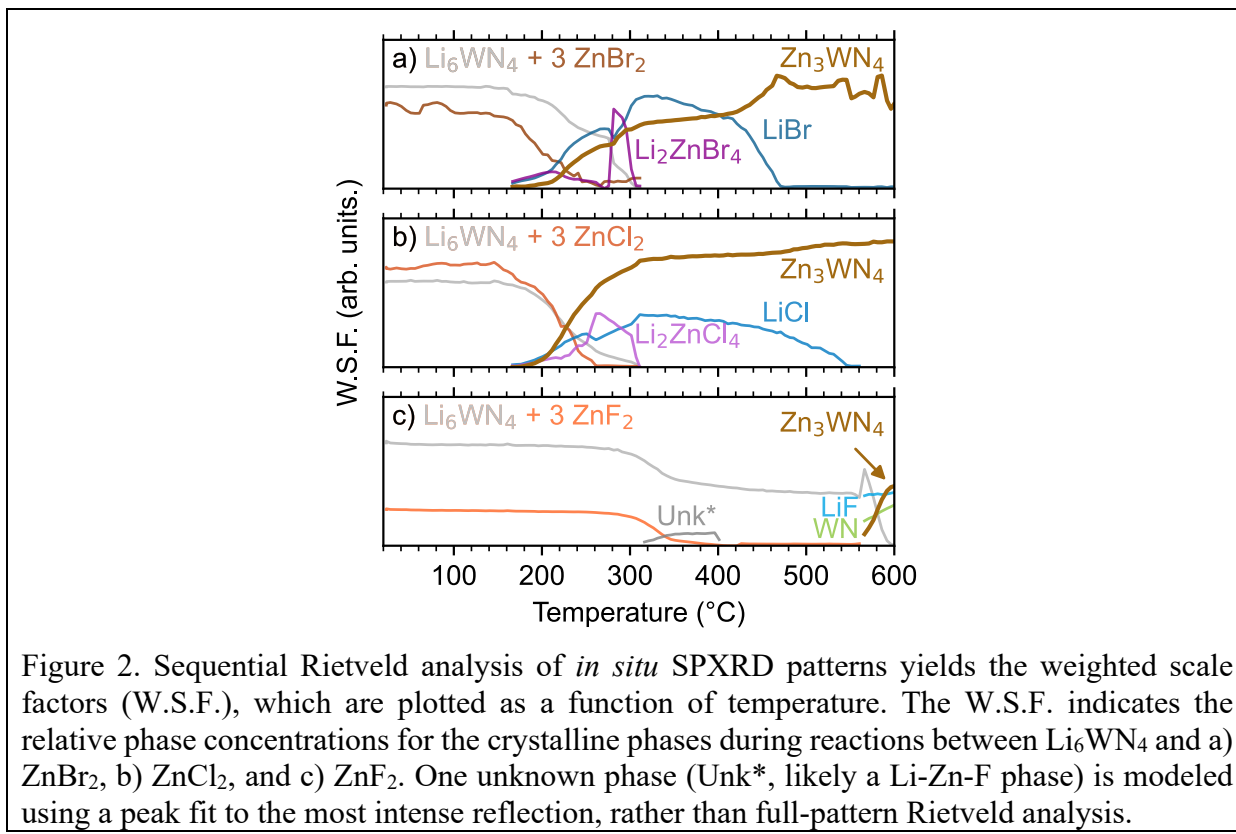

Figure 2. Sequential Rietveld analysis of *in situ* SPXRD patterns yields the weighted scale factors (W.S.F.), which are plotted as a function of temperature. The W.S.F. indicates the relative phase concentrations for the crystalline phases during reactions between Li$_6$WN$_4$ and a) ZnBr$_2$, b) ZnCl$_2$, and c) ZnF$_2$. One unknown phase (Unk*, likely a Li-Zn-F phase) is modeled using a peak fit to the most intense reflection, rather than full-pattern Rietveld analysis.

Sequential Rietveld analysis of *in situ* variable temperature SPXRD measurements of the Li$_6$WN$_4$ + 3 Zn$X_2$ reactions shows that the ZnBr$_2$ and ZnCl$_2$ reactions initiate at much lower temperatures than the ZnF$_2$ reaction (Figure 2). For both the ZnBr$_2$ and ZnCl$_2$ reactions (Figure 2a,b), the concentrations of the precursor phases start decreasing near 170 °C, followed shortly thereafter by Zn$_3$WN$_4$ and Li$X$ formation and growth. Ternary halides Li$_2$ZnBr$_4$ and Li$_2$ZnCl$_4$ are short-lived, incidental intermediates. In contrast, in the fluoride reaction, the concentration of ZnF$_2$ does not begin declining until approximately 300 °C (Figure 2c). The concomitant decrease in Li$_6$WN$_4$ concentration suggests reactivity, but neither Zn$_3$WN$_4$ nor LiF are detected in our data. Instead, very weak reflections for an unknown phase appear in the data (labeled as Unk*). This phase may be a Li-Zn-F ternary, but it does not index to any known ternary fluoride unit cells, including the reported Li$_2$ZnF$_4$ phase.[60] An amorphous phase is likely present in the 400 to 570 °C region, given the decrease in precursor peaks and lack of new intermediate peaks. $Zn_3WN_4$ and LiF crystallize above 570 °C, along with a rocksalt phase (fit with WN, but the material may be a $(Zn,W)N_x$ phase as observed with the Mo-based system, Figure S11). We did not study the $ZnF_2$ reactions further, given that phase-pure $Zn_3WN_4$ did not crystallize and given the challenge associated with washing away LiF from the product. Instead, we focus on $ZnBr_2$ and $ZnCl_2$ reactions.

## DSC measurements

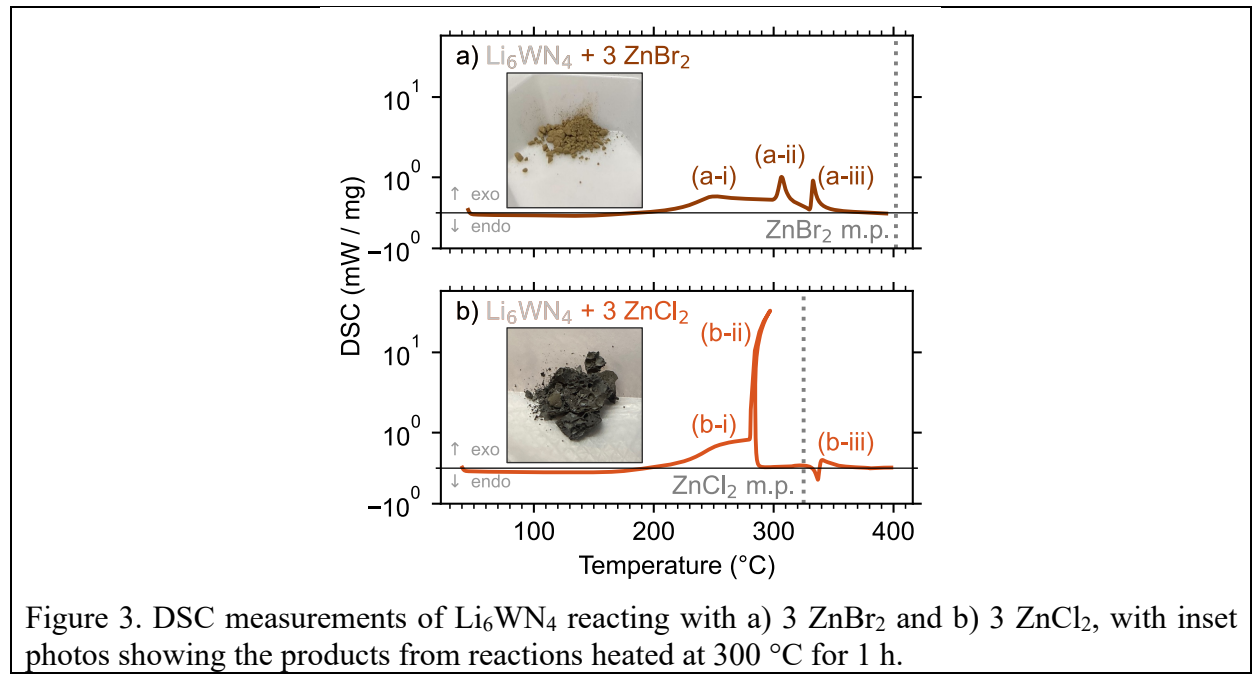

Figure 3. DSC measurements of $Li_6WN_4$ reacting with a) 3 $ZnBr_2$ and b) 3 $ZnCl_2$, with inset photos showing the products from reactions heated at 300 °C for 1 h.

The lower exothermicity of the $ZnBr_2$-based reaction leads to a more controlled release of heat and greater product purity, compared to the $ZnCl_2$-based reaction. DSC measurements show that the $ZnBr_2$ reaction has three small exotherms (Figure 3a). A gradual exotherm starts near 190 °C (a-i), followed by two exotherms near 305 °C (a-ii) and 334 °C (a-iii). This third event is proceeded by a very minor endotherm, possibly consistent with $Li_2ZnBr_4$ melting. This reactivity occurs well below the 392 °C melting point of $ZnBr_2$, suggesting the process is mostly a solid-state reaction. The $ZnCl_2$ reaction (Figure 3b) starts similarly, with a gradual exotherm between 190 °C and 280 °C (b-i). Then at ~280 °C, a massive exotherm (b-ii) initiates just below the melting point of $ZnCl_2$ (325 °C). This event likely corresponds to the formation of a liquid phase, such as a $LiCl$-$ZnCl_2$ eutectic (287 °C at 91% $ZnCl_2$).[61] Peak b-ii has curvature because this event releases heat so quickly that the DSC stage increases in temperature by approximately 15 °C, after which the DSC pan cools slightly. This kind of rapid exothermic event is common in metathesis reactions; once a liquid phase forms, reaction kinetics accelerate and the heat release self-propagates.[38] Lastly, a small endotherm is observed at 336 °C (b-iii), consistent with the melting of $Li_2ZnCl_4$. These results are broadly consistent with the *in situ* SPXRD results, albeit shifted slightly in temperature owing to differences in experimental configuration.

These DSC results show why the $ZnBr_2$ reaction yields the purest product while the $ZnCl_2$ reaction exhibits a small Zn impurity. The rapid release of heat in the $ZnCl_2$ reaction causes small portions of the material to decompose: $Zn_3WN_4$ → W + 3 Zn + 2 $N_2$ or $Zn_3WN_4$ → WN + 3 Zn + 3/2 $N_2$ (Figure S4). In contrast, the washed product of the ZnBr$_2$ synthesis yielded a PXRD pattern with all Bragg peaks indexed to $Pmn2_1$ Zn$_3$WN$_4$. These differences can easily be seen in the color of the material (see insets, Figure 3), where Zn impurities in the ZnCl$_2$ reaction led to a grey color. The Zn$_3$WN$_4$ sample produced by the ZnBr$_2$ reaction is brown. These *in situ* SPXRD and DSC measurements guided our optimization of the synthesis for Zn$_3$WN$_4$.

### Structural and composition analysis of Zn$_3$WN$_4$

The best conditions we found were to heat ZnBr$_2$ with Li$_6$WN$_4$ at a ramp of +5°C/min to 300 °C for a 1 h dwell, followed by natural cooling in the furnace. This reaction was scaled up to ca. 1 g reactant mix for *ex situ* analysis. Washing with anhydrous methanol successfully removed byproduct LiBr and excess ZnBr$_2$ while preserving the targeted phase. We used a slight excess of ZnBr$_2$ (3.1 ZnBr$_2$ + Li$_6$WN$_4$) to ensure complete conversion and minimize reaction temperature by acting as a heat sink. XRF measurements show a Zn:W ratio of 3.165(3):1, slightly higher than the expected 3:1 ratio of Zn$_3$WN$_4$ (representative XRF spectrum shown in Figure S5), which may be a result of the excess ZnBr$_2$. PXRD techniques confirmed that this synthesis of Zn$_3$WN$_4$ proceeded without the formation of decomposition products (i.e., Zn, (Zn,W)N$_x$ phases).

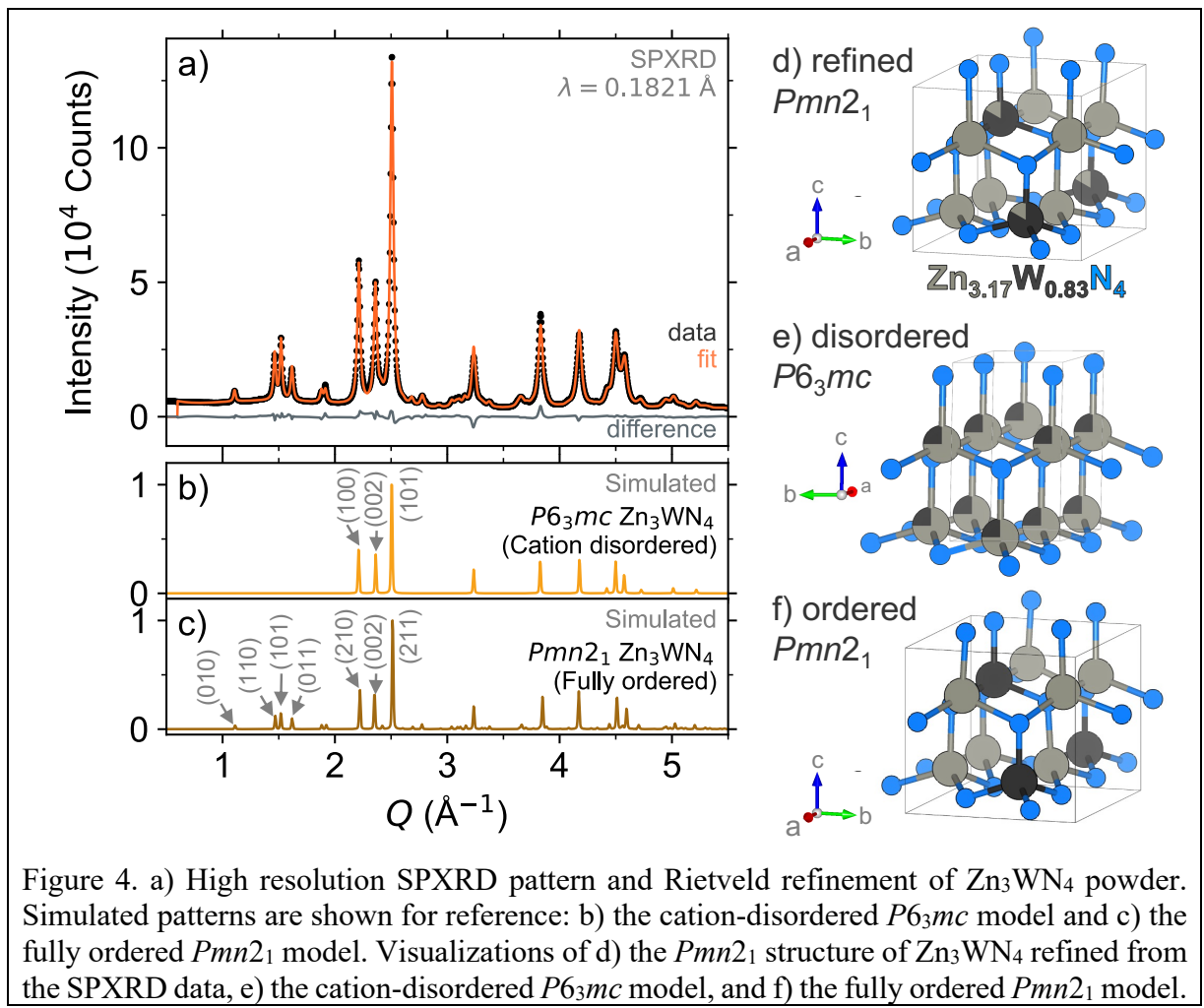

Figure 4. a) High resolution SPXRD pattern and Rietveld refinement of Zn$_3$WN$_4$ powder. Simulated patterns are shown for reference: b) the cation-disordered $P6_3mc$ model and c) the fully ordered $Pmn2_1$ model. Visualizations of d) the $Pmn2_1$ structure of Zn$_3$WN$_4$ refined from the SPXRD data, e) the cation-disordered $P6_3mc$ model, and f) the fully ordered $Pmn2_1$ model.

Rom, *et al.,* 2024    7

High resolution SPXRD measurements confirm the successful synthesis of $Zn_3WN_4$ (Figure 4). Rietveld analysis of the SPXRD data (Figure 4a) shows that $Zn_3WN_4$ crystallizes in space group *Pmn*$2_1$ with lattice parameters $a$ = 6.5602(8) Å, $b$ = 5.6813(7) Å, and $c$ = 5.3235(2) Å. The presence of intensity at the (010), (110), (101), and (011) Bragg positions indicates a substantial degree of cation ordering. The peaks for the (210), (002), and (211) reflections are characteristic of wurtzite-derived structures; these correspond to the (100), (002), and (101) reflections in the prototypical wurtzite structure (*P*$6_3$*mc*), respectively (Figure 4b). Rietveld-refined occupancies suggest a Zn:W ratio of 3.8:1, a higher ratio than that measured by XRF (3.2:1), with partial occupancy of Zn on the W site (Table S1), indicating a composition of $Zn_{3.17}W_{0.83}N_4$ (Figure 4d). The occupancies of the N atoms refined to 1 within error and so were fixed at unity. Alternative structural models were also considered (Table S2, Figures S6–S8), as discussed further in the section on cation ordering in $Zn_3WN_4$.

Property measurements

Diffuse reflectance spectroscopy measurements reveal two absorption onsets for $Zn_3WN_4$: one near 2.5 eV and another near 4.0 eV (Figure 5). This first feature is similar to the 2.0 eV to 2.4 eV absorption onset reported for cation disordered $Zn_3MoN_4$ and $Zn_3WN_4$ synthesized as thin films.[31,62] The second absorption feature occurs near 4.0 eV, which is consistent with the expected bandgap for fully cation-ordered $Zn_3WN_4$ in the *Pmn*$2_1$ space group. The GW-calculated indirect band gap is 3.96 eV, with a direct bandgap of 4.20 eV (NREL MatDB ID 287103; blue trace in Figure 5).[63,64] Other researchers using a hybrid functional, HSE06, calculated the bandgap to be 3.60 eV.[28] These two features therefore suggest that our $Zn_3WN_4$ powder sample is a mixture of cation-ordered and cation-disordered structures. Magnetic susceptibility measurements are consistent with the presence of an impurity, as the material does not exhibit purely diamagnetic behavior (Figure S9).

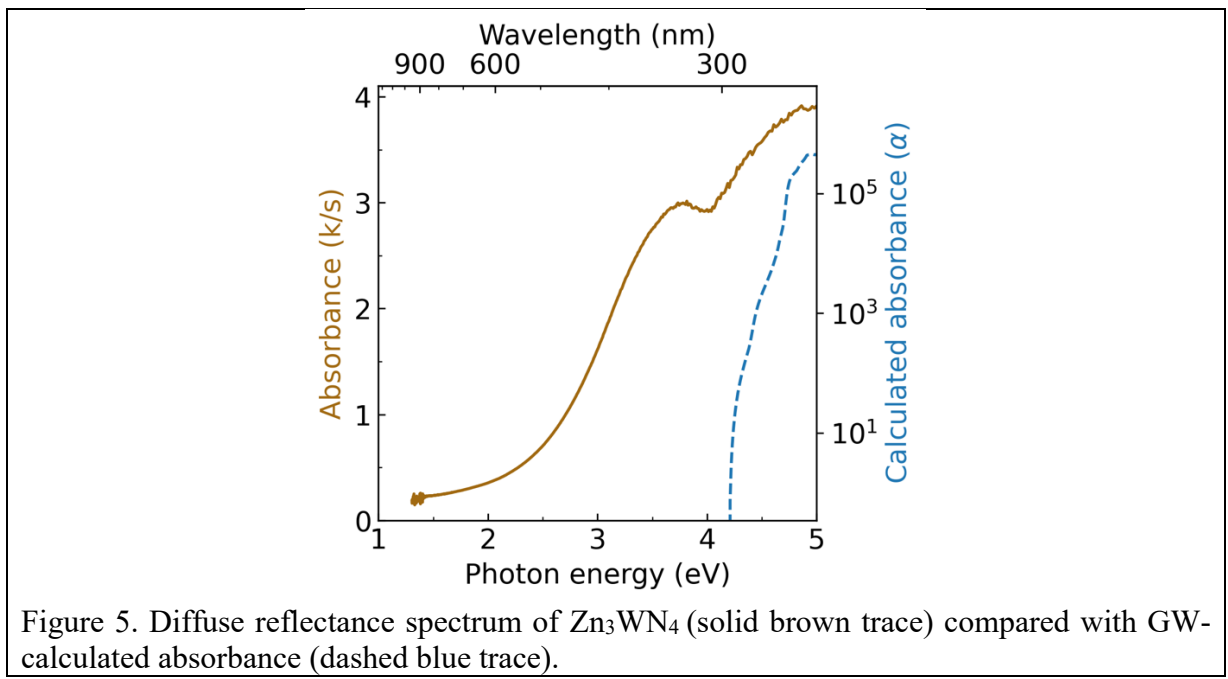

Figure 5. Diffuse reflectance spectrum of $Zn_3WN_4$ (solid brown trace) compared with GW-calculated absorbance (dashed blue trace).

## Cation ordering in $Zn_3WN_4$

Our metathesis approach yielded a different polytype for $Zn_3WN_4$ compared to prior thin film syntheses. We show here that metathesis between $Li_6WN_4$ and $ZnBr_2$ successfully synthesized $Zn_3WN_4$ in space group $Pmn2_1$ (Figure 4). In contrast, prior thin film sputtering work produced cation-disordered $P6_3mc$ structures.[2,31,62] While both the $Pmn2_1$ and $P6_3mc$ structures are wurtzite-derived, the cation-ordered structure is expected to be the thermodynamic ground state.[2,27] In thin film sputtering, high-energy plasma precursors deposit onto a substrate and quench rapidly in a local energy minimum, thus locking in the disordered cation arrangement.[27] While bulk syntheses can sometimes lead to cation-disordered structures,[7,11,47,48,51] the high charge on W (6+) likely encourages ordering to maximize the spacing between the hexavalent cations. The kinetics of our bulk metathesis reactions here proceed in a way that avoids the local energy minimum of the disordered structure, instead forming a (mostly) ordered structure.

The degree of cation ordering in our $Zn_3WN_4$ sample cannot be precisely determined from our current SPXRD data. While the SPXRD results suggest that $Zn_3WN_4$ was synthesized in a (mostly) cation ordered form (Figure 4), our optical spectroscopy results suggest that some degree of cation disorder may be present (Figure 5). Notably, the same batch of $Zn_3WN_4$ was used to create samples for the diffuse reflectance optical spectroscopy, magnetometry, XRF, and high resolution SPXRD measurements. The single-phase model includes some site disorder, leading to a refined composition of $Zn_{3.17}W_{0.83}N_4$ and a good statistical fit to the SPXRD pattern ($R_{wp}$ = 3.989%). Attempts to model the pattern with two phases (a cation-ordered $Pmn2_1$ and a cation-disordered $P6_3mc$) yield similar-quality fits as the single-phase fit (Figure S6). The two-phase refinements suggest the powder may be approximately 10–20 mol% $P6_3mc$. These two-phase models are consistent with our optical data, as disordered $P6_3mc$ $Zn_3WN_4$ likely exhibits lower-energy absorption (ca. 2.5 eV)[31] compared to cation-ordered $Pmn2_1$ $Zn_3WN_4$, which has a predicted bandgap of ca. 4.0 eV.[28] However, the small difference in statistical fit for the best two-phase model ($R_{wp}$ = 3.742%) relative to the best single-phase model ($R_{wp}$ = 3.989%) suggests that determining the precise cation order of the material is non-trivial (see the Supplemental Information for more discussion). In sum, these results indicate that our batch of cation-ordered $Zn_3WN_4$ exhibits some degree of disorder or inhomogeneity on the order of 10–20 mol%. Although our $Zn_3WN_4$ sample is not perfectly ordered, it exhibits a substantially larger degree of cation ordering than prior thin film work.[31,62]

## Structural relation of precursor and product

The synthesis reported here is distinct from literature on prior ion exchange syntheses of nitrides in that the ions undergoing exchange have different formal charges. All prior reports on nitrides exchanged ions of the same charge (e.g., displacing $Na^+$ with $Cu^+$ in $ATaN_2$, or $Ca^{2+}$ with $Mg^{2+}$ in $A_2Si_5N_8$; where $A$ represents the exchangeable cations).[49,52,53] Here, we replace a monovalent cation ($Li^+$) with a divalent cation ($Zn^{2+}$). While such exchange has been conducted in oxides[65,66] and sulfides (e.g., 2 $NaCrS_2$ + $MgCl_2$ → $MgCr_2S_4$ + 2 NaCl),[67] to the best of our knowledge this is the first report of such an exchange in nitrides. The resulting decrease in the cation:anion ratio (from 7:4 to 4:4) means that a truly topotactic replacement is unlikely to occur. However, the transformation appears to be near-topotactic.

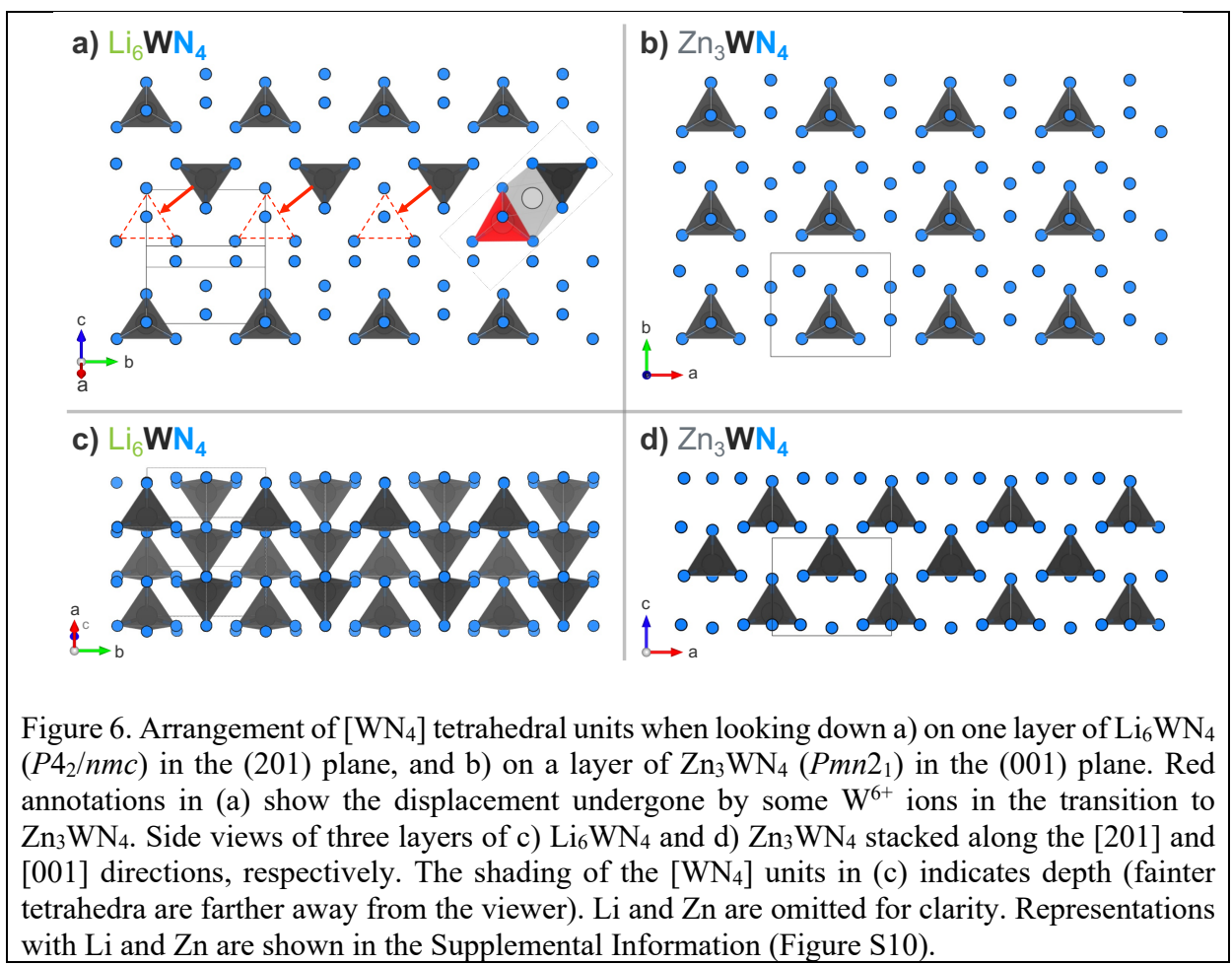

Figure 6. Arrangement of [WN$_4$] tetrahedral units when looking down a) on one layer of Li$_6$WN$_4$ (*P*4$_2$/*nmc*) in the (201) plane, and b) on a layer of Zn$_3$WN$_4$ (*Pmn*2$_1$) in the (001) plane. Red annotations in (a) show the displacement undergone by some W$^{6+}$ ions in the transition to Zn$_3$WN$_4$. Side views of three layers of c) Li$_6$WN$_4$ and d) Zn$_3$WN$_4$ stacked along the [201] and [001] directions, respectively. The shading of the [WN$_4$] units in (c) indicates depth (fainter tetrahedra are farther away from the viewer). Li and Zn are omitted for clarity. Representations with Li and Zn are shown in the Supplemental Information (Figure S10).

The transformation from Li$_6$WN$_4$ to Zn$_3$WN$_4$ involves slight structural rearrangements (Figure 6). The W$^{6+}$ retains its tetrahedral coordination and (for the most part) its oxidation state through the process, but the orientations of the polyhedra change. The fcc anion lattice of Li$_6$WN$_4$ converts to the hcp anion lattice of Zn$_3$WN$_4$. During this change, half of the W$^{6+}$ ions in Li$_6$WN$_4$ migrate through an octahedral site to a new tetrahedral site in Zn$_3$WN$_4$ (red annotations). The anion packing layers also decrease in spacing from 2.767(1) Å in Li$_6$WN$_4$ to 2.662(1) Å in Zn$_3$WN$_4$. The shortest W–W distance decreases from 4.927(1) Å to 4.646(1) Å. Lastly, the centrosymmetric structure of Li$_6$WN$_4$ (*P*4$_2$/*nmc*) converts to a polar structure of Zn$_3$WN$_4$ (*Pmn*2$_1$). We did not observe any signs of solid solution behavior (i.e., Li$_{6-x}$Zn$_{x/2}$WN$_4$) in the *in situ* SPXRD studies. However, solid solution behavior may be present but undetected by the *in situ* SPXRD data if it occurs on short timescales (<30 s) or small length scales (ca 10 nm).

### Reaction pathway

Our results demonstrate that the halide anion *X* in Zn*X*$_2$ exerts a powerful influence over the metathesis reaction thermodynamics and kinetics of Zn-containing ternary nitrides. *In situ* SPXRD and DSC measurements reveal that the phase purity of the product is correlated with the melting points of the phases and the reaction energy (Figure 7). The ZnF$_2$-based reaction is highly exothermic, and ZnF$_2$ has a high melting point. These two factors result in high-temperature reactivity and partial Zn$_3$WN$_4$ phase decomposition. In contrast, ZnCl$_2$ and ZnBr$_2$ react at much

Rom, *et al.,* 2024    10

lower temperatures (owing to the formation of liquid phases) and release less energy during the reaction. However, the slightly lower melting point of $ZnCl_2$ compared to $ZnBr_2$ combined with the slightly higher $\Delta H_{rxn}$ of the respective reaction leads to a runaway exothermic event (Figure 3) and slight $Zn_3WN_4$ decomposition (Figure S4). Overall, the $ZnBr_2$-based reaction is optimal for the synthesis of $Zn_3WN_4$ because of the low $\Delta H_{rxn}$ and the low melting point of $ZnBr_2$, but further optimization for phase purity may be needed in future research on other materials.

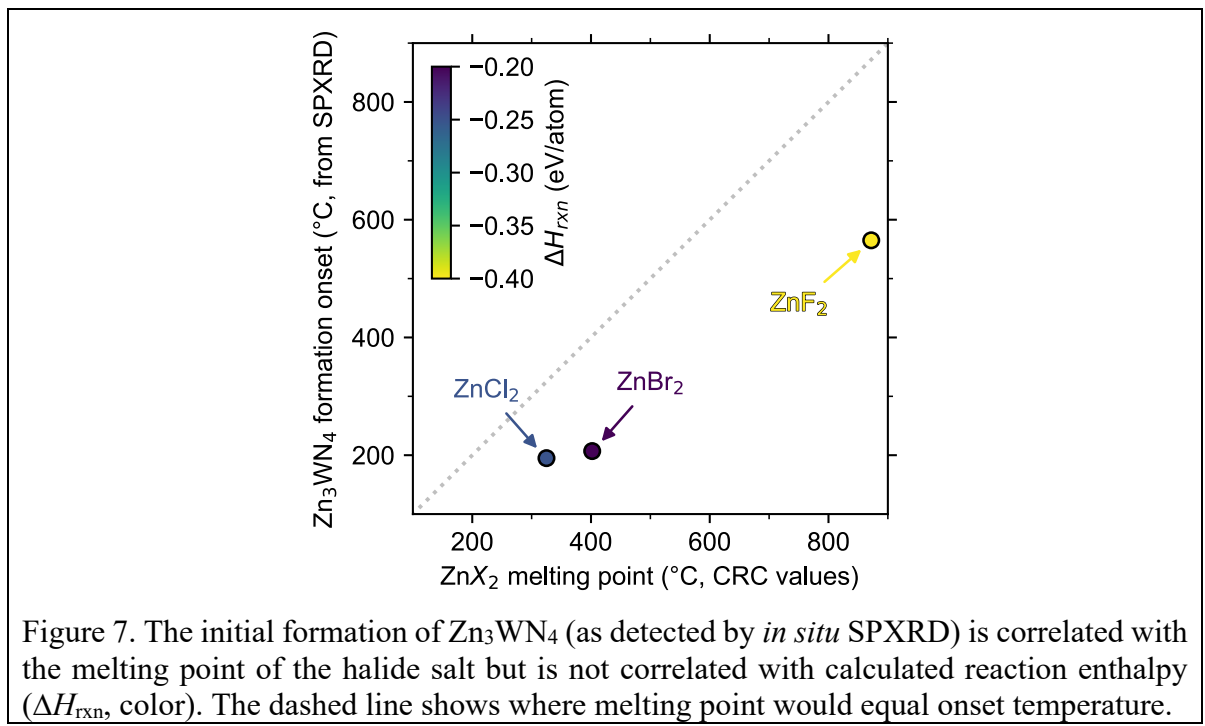

Figure 7. The initial formation of $Zn_3WN_4$ (as detected by *in situ* SPXRD) is correlated with the melting point of the halide salt but is not correlated with calculated reaction enthalpy ($\Delta H_{rxn}$, color). The dashed line shows where melting point would equal onset temperature.

This type of reaction control has been explored in oxides but has not previously been detailed for ternary nitrides. "Spectator ions" that are not incorporated into the final product still have substantial influence over reaction pathways and polymorph formation, as demonstrated for syntheses of Y-Mn-O phases.[65,66,68–71] In particular, work on "co-metathesis" identified that when eutectic halide mixtures form *in situ*, these liquids decrease reactant onset temperatures relative to systems without eutectics.[65,71] Similar eutectics are likely forming between $ZnX_2$ and $LiX$ in our syntheses of $Zn_3WN_4$.

Generalizability to other materials
There are numerous Zn-*M*-N phases that have been demonstrated to be synthesizable via thin film sputtering but that have not yet been made in bulk. In addition to $Zn_3WN_4$,[2,62] sputtering has been used to synthesize fully nitridized transition metal ternaries: $ZnTiN_2$, $ZnZrN_2$, $Zn_2VN_3$, $Zn_2NbN_3$, $Zn_2TaN_3$, and $Zn_3MoN_4$.[2,3,26,27,31] Although computational predictions for these thin film materials find that cation-ordered structures are the thermodynamic ground state (Figure 4f),[72] these sputtered films tend to form in cation-disordered structure variants (Figure 4e).[2,3] This disorder tends to decrease the bandgap of the material by creating localized electronic states.[8,73] Bulk syntheses of these materials could advance the development of these new semiconductors by studying the effect of structure (e.g., ordering) on optoelectronic properties of these new materials.

The synthesis of $Zn_3WN_4$ from $Li_6WN_4$ and $ZnX_2$ suggests a promising strategy for future materials discovery of cation-ordered heterovalent ternary nitrides via metathesis from lithium-based ternary nitride precursors. Lithium-based ternary nitrides are the most well-studied subset of ternary nitrides,[2] suggesting that many Li-$M$-N phases exist that could be used to synthesize additional $A$-$M$-N phases via exchange with $AX_n$ (where $A$ and $M$ are metals and $X$ is a halide). Following our results here, $X$ should be selected to minimize reaction energy and thus minimize the risk of decomposing the target phase via gaseous $N_2$ loss. To demonstrate this point, we also synthesized $Zn_3MoN_4$ from $Li_6MoN_4$ and $ZnBr_2$ (Figure S11). $Zn_3MoN_4$ was the main product, but some decomposition products were also observed, indicating that additional reaction optimization is needed. Unlike in $Zn_3WN_4$, the $ZnBr_2$-based reaction was not sufficiently low-energy to avoid this decomposition for $Zn_3MoN_4$. While we were not able to synthesize phase-pure $Zn_3MoN_4$ here, further reaction engineering, like adding $NH_4Cl$ to manage heat flow,[42,43,45,46] may be able to produce phase-pure $Zn_3MoN_4$. As we found that the reaction onset temperature is correlated with $AX_2$ melting point, phases with high-melting temperature precursors may be difficult to synthesize below the decomposition point of the targeted ternary. Therefore, future work should consider ways to decouple the reaction onset from the $AX_2$ melting point. In sum, this work shows how Li-$M$-N phases can be promising precursors for accelerating the discovery of new ternary nitrides.

## Conclusions

Here, we report the bulk synthesis of cation-ordered $Zn_3WN_4$, through ion exchange reactions beginning from Li-based ternary nitride precursors: $Li_6WN_4 + ZnX_2 \rightarrow Zn_3WN_4 + 6\ LiX$ ($X$ = Br, Cl, F). These reactions proceed directly (i.e., without intermediates), as measured by *in situ* synchrotron powder X-ray diffraction and differential scanning calorimetry. The reaction onset temperature correlates with the melting point of the $ZnX_2$ precursor, allowing $ZnCl_2$- and $ZnBr_2$-based reactions to proceed at ≤300 °C. The more exothermic reactions lead to greater degrees of $Zn_3WN_4$ decomposition, meaning that the least exothermic reaction (with $ZnBr_2$) is the most favorable for synthesis. High resolution synchrotron powder X-ray diffraction data is consistent with cation-ordered $Zn_3WN_4$ (*Pmn*$2_1$). This finding is distinct from prior thin film syntheses, which yielded cation-disordered *P*6$_3$*mc* $Zn_3WN_4$. Diffuse reflectance spectroscopy shows that $Zn_3WN_4$ powders exhibit absorption onsets near 2.5 eV and 4.0 eV, suggesting that a small amount of a cation-disordered *P*6$_3$*mc* phase of $Zn_3WN_4$ may be mixed in with the cation-ordered *Pmn*$2_1$ phase. Preliminary work targeting $Zn_3MoN_4$ from $Li_6MoN_4$ and $ZnBr_2$ suggests this synthesis approach may readily extend to other systems. These findings indicates that Li-$M$-N compounds may serve as precursors for synthesizing numerous other ternary nitrides.

## Acknowledgements

This work was authored at the National Renewable Energy Laboratory, operated by Alliance for Sustainable Energy, LLC, for the U.S. Department of Energy (DOE) under Contract No. DE-AC36-08GO28308. Funding provided by DOE Basic Energy Sciences Early Career Award "Kinetic Synthesis of Metastable Nitrides." This research used resources of the Advanced Photon Source, a U.S. DOE Office of Science user facility operated for the DOE Office of Science by Argonne National Laboratory under Contract No. DE-AC02-06CH11357. R.A.K. gratefully acknowledges support from the U.S. DOE Office of Energy Efficiency and Renewable Energy (EERE), Hydrogen and Fuel Cell Technologies Office (HFTO). R.W.S. acknowledges support from the Director's Fellowship within NREL's Laboratory Directed Research and Development program, and K.H. acknowledges support from the DOE Science Undergraduate Laboratory Internships (SULI) program. The authors thank Dr. Robert Bell and Dr. Noemi Leick for technical support with DSC measurements, Dr. Fred Baddour for support with the laboratory PXRD measurements, Dr. Wenqian Xu for support with the *in situ* SPXRD measurements, and Dr. Stephan Lany for calculating the absorption coefficient of *Pmn*$2_1$ Zn$_3$WN$_4$. The views expressed in the article do not necessarily represent the views of the DOE or the U.S. Government.

## Author Contributions

Christopher L. Rom: Conceptualization, Investigation, Formal analysis, Visualization, Writing-Original draft preparation
Shaun O'Donnell: Investigation
Kayla Huang: Investigation
Ryan A. Klein: Investigation, Formal analysis
Morgan J. Kramer: Investigation, Formal analysis
Rebecca W. Smaha: Investigation, Writing-Reviewing and editing
Andriy Zakutayev: Conceptualization, Funding acquisition, Supervision, Writing-Reviewing and editing

# Supplemental Information for:
# Bulk synthesis of Zn$_3$WN$_4$ via solid-state metathesis

Christopher L. Rom[1], Shaun O'Donnell[1,2], Kayla Huang[1,3], Ryan A. Klein[1,4], Morgan J. Kramer[4,5], Rebecca W. Smaha[1], Andriy Zakutayev[1,*]

[1] Materials, Chemical, and Computational Science, National Renewable Energy Laboratory, Golden, CO, 80401, USA
[2] Department of Chemistry, Colorado State University, Fort Collins, CO, 80523, USA
[3] University of Illinois Urbana-Champaign, Champaign, IL, 61801, USA
[4] Center for Neutron Research, National Institute of Standards and Technology, Gaithersburg, MD, 20899, USA
[5] Department of Chemistry, Southern Methodist University, Dallas, TX, 75275 USA

[*] Corresponding author: andriy.zakutayev@nrel.gov

## Table of Contents

## Experimental

### Synthesis of $Zn_3WN_4$

As some precursors are highly moisture sensitive, all precursors were prepared and stored in an argon-filled glovebox ($O_2 < 0.1$ ppm, $H_2O < 0.1$ ppm) unless explicitly mentioned. $ZnF_2$ (≥99 %, anhydrous, Alfa Aesar), $ZnCl_2$ (≥99.995%, anhydrous, Sigma Aldrich), $ZnBr_2$ (99.999%, anhydrous, Aldrich), $Li_3N$ (≥99.5%, 80 mesh, Sigma Aldrich), W (99.95%, <1 micron powder, Thermofisher Scientific) were used as received.

$Li_6WN_4$ was synthesized in a method modified from that of Yuan et al.[1] Solid precursors (2.1 $Li_3N$ + W, ca. 5 mol% excess $Li_3N$ to account for loss by evaporation) were ground with a mortar and pestle and loaded into Zr crucibles with Zr lids (ca. 1 g loose powder). The Zr crucibles were then loaded into sacrificial quartz tubes (open on one end), which were loaded into quartz process tubes and heated in a tube furnace. Custom endcaps with quick disconnects enabled air-free transfer from the glovebox to the tube furnace (under Ar or $N_2$). The samples were reacted under flowing $N_2$ (50 sccm, 99.999% purity) with a +5 °C/min ramp followed by a 12 h dwell at 850 °C and then natural cooling after turning off the furnace. Samples were recovered into the glovebox for subsequent analysis and use. The beige-colored products were confirmed to be phase pure by powder X-ray diffraction (Figure S1).

Syntheses for $Zn_3WN_4$ were conducted by grinding together $Li_6WN_4$ with $ZnX_2$ ($X$ = Cl, Br) in a ratio of approximately 1:3. The powders were pelletized with 6 mm diameter dies in an arbor press (ca. 100 mg per pellet), loaded into quartz ampules (10 mm OD, 10 mm ID, ca 10 cm length), sealed under vacuum (<0.03 Torr), and heated in a muffle furnace. The optimized synthesis for $Zn_3WN_4$ used a ratio of $Li_6WN_4$ + 3.1 $ZnBr_2$ and was scaled up to a 3 g batch, sealed in a quartz ampule under vacuum (< 0.03 Torr) and heated at +5 °C/min to 300 °C for a 1 h dwell, then allowed to cool naturally. Samples were recovered into the glovebox. Reaction byproduct $LiX$ was washed away using anhydrous and degassed methanol that was dried over molecular sieves. For washing, centrifuge tubes were loaded with approximately 500 mg of product powder and 1.5 mL methanol. The tube was agitated with a vortex, centrifuged, and the supernatant was decanted. This wash was repeated for a total of 3 cycles. Recovered powders were dried overnight under vacuum. However, $Zn_3WN_4$ ultimately proved to be stable against air and water, and we note that the anhydrous washing may not be necessary.

### Synthesis of $Zn_3MoN_4$

Just like $Li_6WN_4$, $Li_6MoN_4$ was synthesized using $Li_3N$ and Mo (≥99.9%, 1–5 micron powder, Sigma Aldrich), following a method modified from that of Yuan et al.[1] Heating the powders at 850 °C for 12 h (with slight $Li_3N$ excess) resulted in a phase pure $Li_6MoN_4$ (Figure S1b). This $Li_6MoN_4$ was then mixed with $ZnBr_2$, pelletized, sealed in an ampule under vacuum, and heated at +5 °C/min to 300 °C for a 1 h dwell, followed by natural cooling. The product was then washed with anhydrous methanol.

*In situ* synchrotron powder X-ray diffraction analysis

*In situ* synchrotron powder X-ray diffraction (SPXRD) measurements were conducted at beamline 17-BM-B of the Advanced Photon Source at Argonne National Laboratory. For these experiments (λ = 0.24101 Å), the PerkinElmer plate detector was positioned 700 mm away from the sample. Homogenized precursors were packed into quartz capillaries in an Ar glovebox and flame-sealed under vacuum (<30 mTorr). Capillaries were loaded into a flow-cell apparatus[3] and heated at 5 °C/min to the specified temperature. A thermocouple was placed against the tip of the sample capillary, approximately 2 mm horizontally from the position of the X-ray beam. Diffraction pattern images were collected every 30 s by summing 20 exposures of 0.5 s each (10 s of summed exposure), followed by 20 s of deadtime. Images collected from the plate detector were radially integrated using GSAS-II and calibrated using a silicon standard.

Sequential Rietveld refinements were conducted on *in situ* SXPRD datasets using TOPAS Professional v6.[4] Lattice parameters, background terms, and scale factors were refined for each phase as a function of temperature, while atomic coordinates and occupancies were held constant at the initial values of the reference structure. A weighted scale factor (W.S.F.) $Q$ was calculated for each phase $p$ as a product of scale factor $S$, cell volume $V$, and cell mass $M$: $Q_p = S_p \cdot V_p \cdot W_p$.[5] We note that amorphous and liquid phases are inherently not observed in powder diffraction measurements and therefore cannot be accurately included in this analysis. A Lorentzian size broadening term was refined for each phase to model the peak shape using the pattern showing the greatest intensity of the relevant phase; this term was then fixed for the sequential refinements to better account for changes in intensity. To help stabilize the sequential refinement, isotropic displacement parameters ($B_{iso}$) were fixed at 1 Å$^2$ for all atoms, but we note that this is likely not physical for a variable temperature investigation.

*Ex situ* powder X-ray diffraction analysis of $Zn_3WN_4$

The products of all reactions were characterized by powder X-ray diffraction (PXRD). Laboratory X-ray diffraction patterns were collected on a Rigaku Ultima IV diffractometer with Cu Kα X-ray radiation at room temperature. All samples were initially prepared for PXRD measurements inside the glovebox; powder was placed on off-axis cut silicon single crystal wafers to reduce background scattering and then covered with polyimide tape to impede exposure to atmosphere. After $Zn_3WN_4$ was determined to be moderately air stable, PXRD patterns were collected without polyimide tape to decrease the background signal.

High resolution synchrotron powder X-ray diffraction (SPXRD) measurements were conducted at beamline 28-ID-2 of the National Synchrotron Light Source II (λ = 0.1821 Å) at Brookhaven National Laboratory. Samples were sealed under vacuum in quartz capillaries, which were then nested in Kapton capillaries. Data were collected for 60 seconds at $T$ = 25 °C while spinning. Scattered photon intensity was measured using a Perkin-Elmer XRD 1621 Digital Imaging Detector. The data were reduced using Dioptas.[6] Pawley fits and subsequent Rietveld refinements were conducted using TOPAS Academic.[4]

Rietveld refinements were conducted for the laboratory PXRD and SPXRD patterns using TOPAS and TOPAS Academic, v6 (Bruker AXS).[4] Reference structures were sourced from the Inorganic

Crystal Structure Database (ICSD). The $Pmn2_1$ $Zn_3MoN_4$ structure (ICSD Col. Code 255744) was used as a starting model for $Pmn2_1$ $Zn_3WN_4$, with the Mo replaced by W.[7] For the cation disordered $P6_3mc$ $Zn_3WN_4$ structure, $P6_3mc$ ZnO was used as a starting model, with atomic occupancies adjusted to match the stoichiometry of $Zn_3WN_4$, and lattice parameters adjusted to match the SPXRD pattern. For each structure, lattice parameters, isotropic displacement parameters, and general atomic coordinates were refined. For some models, Zn and W occupancy were refined as detailed in the Discussion section and the Supplemental (Figure S7). Structural visualizations and reference PXRD patterns were generated using VESTA.[8]

Compositional, thermodynamic, and property measurements

The composition of nominal $Zn_3WN_4$ was measured by X-ray Fluorescence spectroscopy (XRF) and combustion analysis. Cation composition was quantified by XRF using a Bruker M4 Tornado with a Rh X-ray source. Samples were pelletized and XRF spectra were collected at 4 points across the pellet. Zn and W ratios were quantified from each spectra using the Bruker M4 software.

Differential scanning calorimetry (DSC) experiments were conducted using a Q20 system from TA Instruments. Samples were prepared in an Ar-filled glovebox. Samples (ca. 10 mg) were loaded into aluminum pans, which were crimped closed with an aluminum lid. The reference pan was also crimped closed under argon. Pans were then transferred out of the glovebox for measurement, and data were collected upon ramping up to 400 °C at a rate of 10 °C/min. Thermodynamic calculations for reaction enthalpies ($\Delta H_{rxn}$) were conducted using formation enthalpy ($\Delta H_f$) values reported in the Materials Project database.[9,10]

UV-vis measurements were conducted on a Cary 6000 UV-Vis-NIR spectrometer. PTFE was used as a white reflectance standard. Absorbance was calculated with the Kubelka-Munk transformation, $k/s = (1 - R)^2 / 2R$ (where $R$ is the reflectance, $k$ is the apparent absorption coefficient, and $s$ is the apparent scattering coefficient).

DC susceptibility data were measured on a Quantum Design Physical Properties Measurement System (PPMS) from $T = 2$ to 305 K in applied fields up to $\mu_0H = 14$ T.

## Additional PXRD measurements

Laboratory PXRD confirms the phase purity of the precursors that were synthesized for our metathesis reactions (Figure S1). All peaks index to the reference structures from the ICSD. The broad background at $2\theta < 35°$ comes from the polyimide tape used to protect the powder from air and moisture. This background suppresses the measured peak intensities for peaks at $2\theta < 35°$, leading to the poor fit for that region of the pattern.

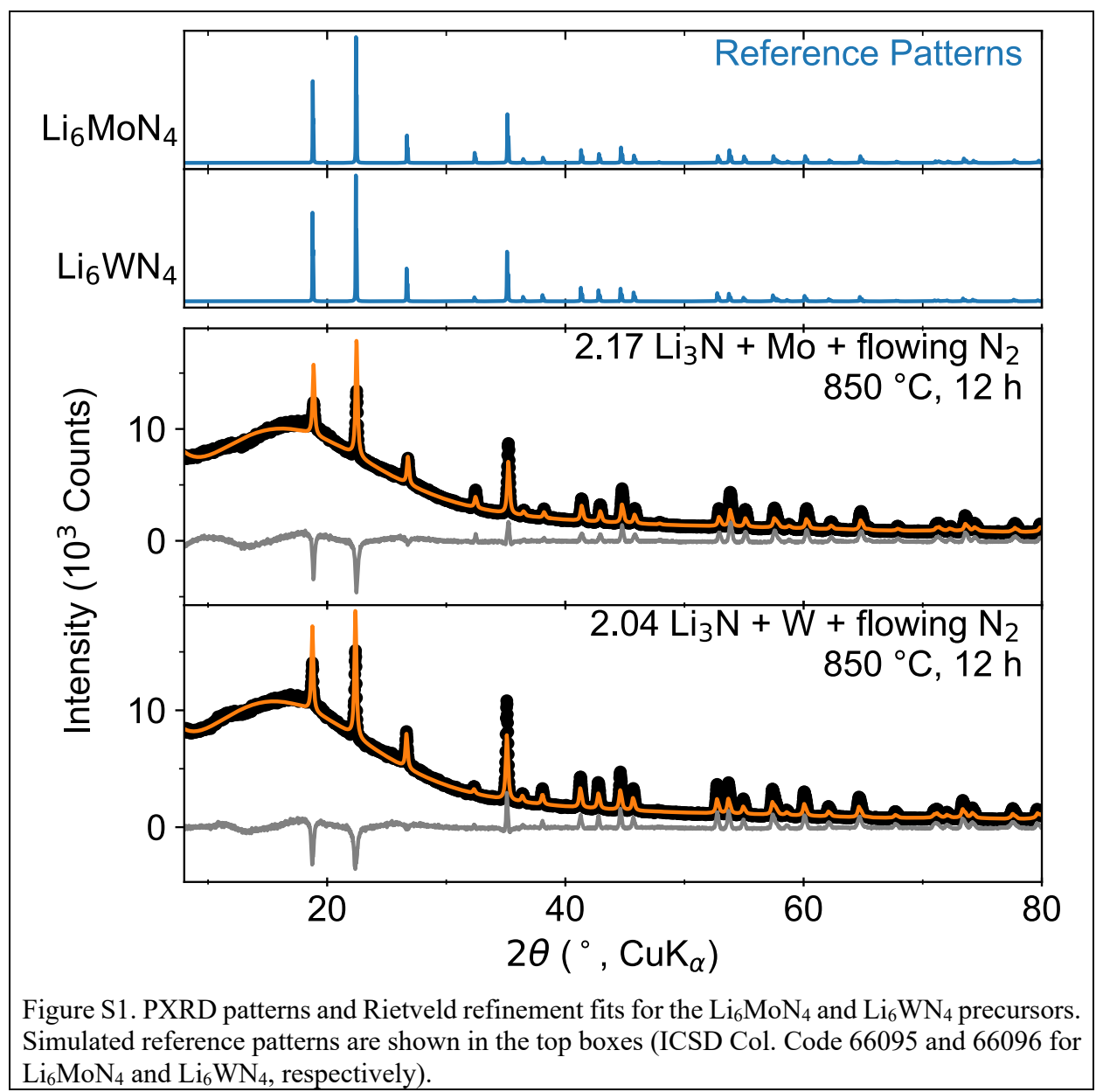

Figure S1. PXRD patterns and Rietveld refinement fits for the $Li_6MoN_4$ and $Li_6WN_4$ precursors. Simulated reference patterns are shown in the top boxes (ICSD Col. Code 66095 and 66096 for $Li_6MoN_4$ and $Li_6WN_4$, respectively).

## *In situ* variable temperature SPXRD measurements

Figures S2 and S3 show the *in situ* variable temperature SPXRD heatmaps for the $ZnCl_2$ and $ZnF_2$ reactions, respectively. The analogous $ZnBr_2$ reaction is shown in the main text (Figure 1). These data were used for sequential Rietveld analysis, which is presented in Figure 4 in the main text.

*In situ* SPXRD of the 3 $ZnCl_2$ + $Li_6WN_4$ reaction (Figure S2) proceeds similarly to the $ZnBr_2$-based reaction shown in the main text (Figure 1). Bragg peaks arising from the precursors (3 $ZnCl_2$ + $Li_6WN_4$) stay steady up to 200 °C, where they begin to decrease in intensity. Simultaneously, Bragg peaks for $Zn_3WN_4$, LiCl, and $Li_2ZnCl_4$ begin to increase in intensity. The set of Bragg peaks corresponding to the $Li_2ZnCl_4$ phase fade out by 310 °C. We suspect that $Li_2ZnCl_4$ is not an essential intermediate but rather the product of a transient side reaction between the precursor $ZnCl_2$ and product LiCl. The intensity of the Bragg peaks corresponding to the LiCl phase reaches a maximum near ca. 340 °C, and then slowly decreases in intensity up to 550 °C (LiCl melting point is 605 °C).[11] $Zn_3WN_4$'s Bragg peaks remain approximately constant in intensity above 300 °C, persisting through the duration of the heating process. These processes are consistent with the following reactions:

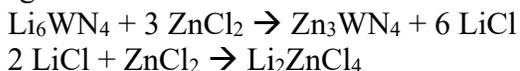

$Li_6WN_4$ + 3 $ZnCl_2$ → $Zn_3WN_4$ + 6 LiCl
2 LiCl + $ZnCl_2$ → $Li_2ZnCl_4$

Figure S3 shows that the 3 $ZnF_2$ + $Li_6WN_4$ reaction does not yield $Zn_3WN_4$ until 565 °C, a much higher temperature than the $ZnCl_2$ and $ZnBr_2$ reactions. Initial reactivity begins near 310 °C, indicated by a decrease in intensities for the set of Bragg peaks corresponding to $Li_6WN_4$ and $ZnF_2$. This initial reactivity is well below the melting point of $ZnF_2$ (872 °C).[11] Concurrent with this initial reaction, an unknown phase briefly grows in (between 310 °C and 404 °C). Extrapolating from the $ZnBr_2$ and $ZnCl_2$ reactions, the phase is likely a Li-Zn-F intermediate, but ternary fluorides in this space are poorly characterized. $Li_2ZnF_4$ has been reported, but the structure is not well described and the unit cell does not match the unknown phase. This intensity of the Bragg peaks arising from the intermediate phase decreases to zero by 404 °C, above which $Li_6WN_4$ is the only crystalline phase up to 565 °C. At this point, $Li_6WN_4$ fades out and $Zn_3WN_4$, LiF, and a rocksalt phase fit with WN grow in. The presence of this rocksalt phase indicates that the higher reaction temperature and greater exothermicity of the $ZnF_2$ reaction (compared with the Cl and Br versions) leads to a substantial degree of decomposition of the $Zn_3WN_4$ phase. Therefore, this reaction was not explored further.

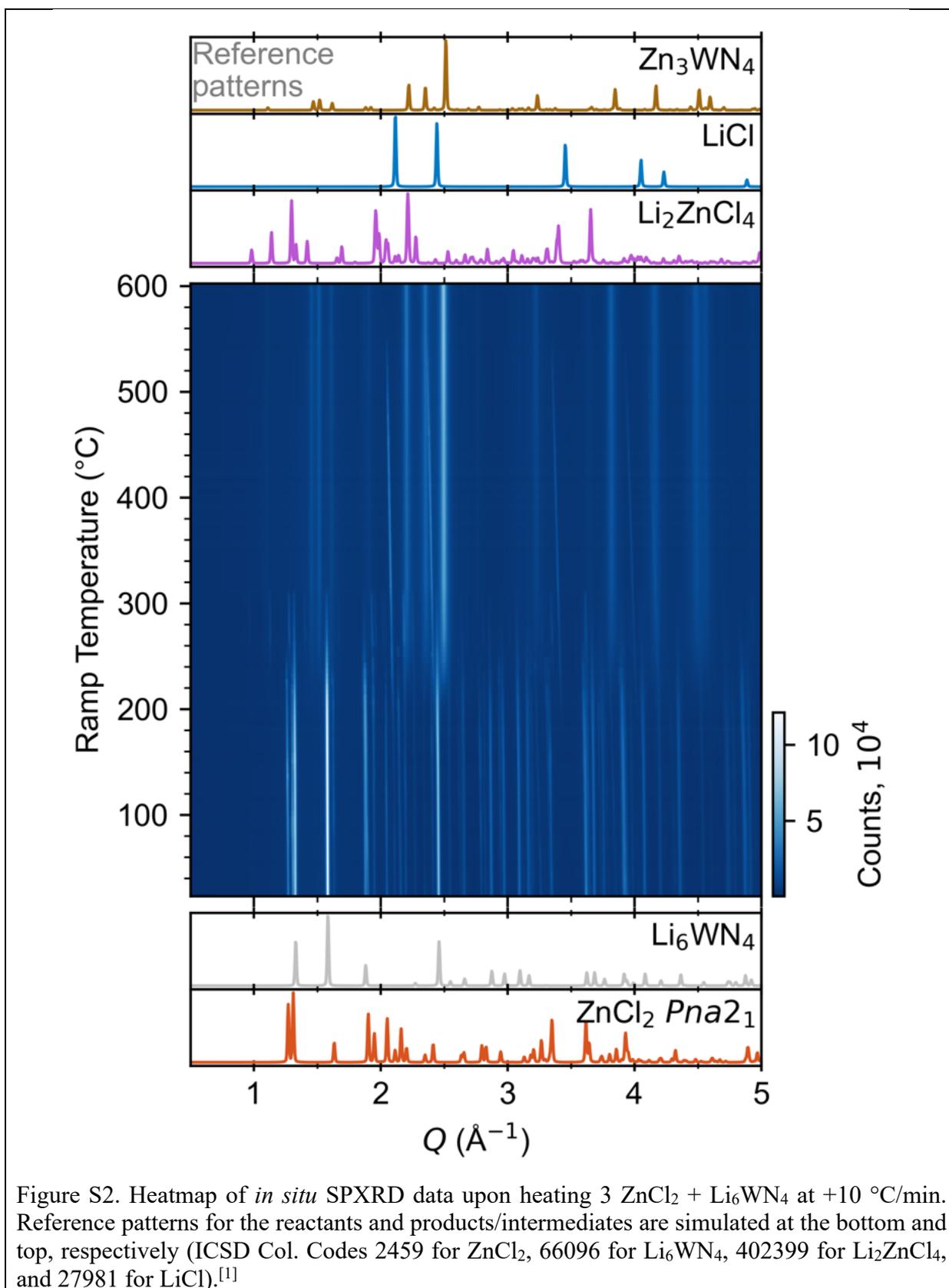

Figure S2. Heatmap of *in situ* SPXRD data upon heating 3 $ZnCl_2$ + $Li_6WN_4$ at +10 °C/min. Reference patterns for the reactants and products/intermediates are simulated at the bottom and top, respectively (ICSD Col. Codes 2459 for $ZnCl_2$, 66096 for $Li_6WN_4$, 402399 for $Li_2ZnCl_4$, and 27981 for LiCl).[1]

Rom, *et al.*, 2024    23

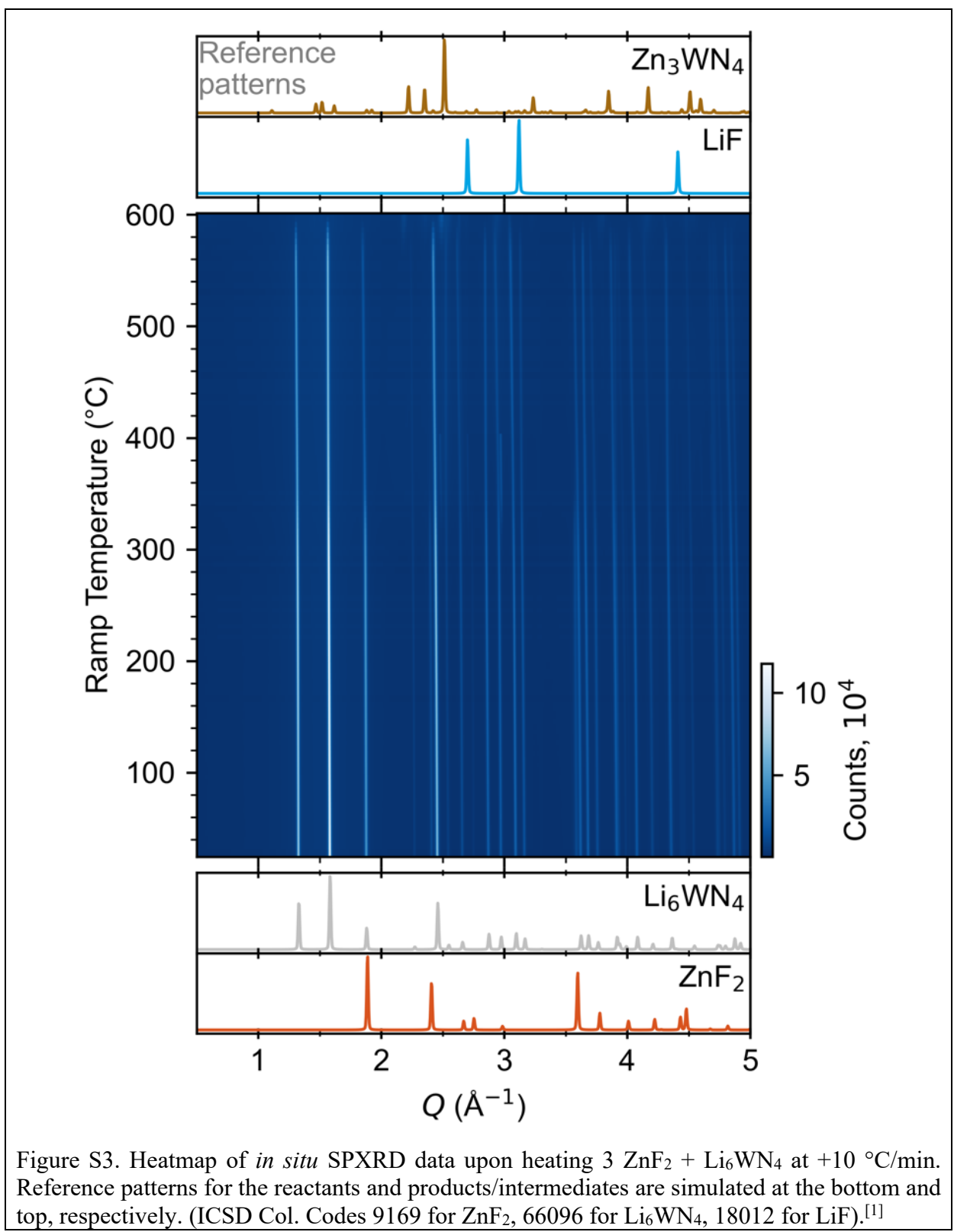

Figure S3. Heatmap of *in situ* SPXRD data upon heating 3 $ZnF_2$ + $Li_6WN_4$ at +10 °C/min. Reference patterns for the reactants and products/intermediates are simulated at the bottom and top, respectively. (ICSD Col. Codes 9169 for $ZnF_2$, 66096 for $Li_6WN_4$, 18012 for LiF).[1]

Rom, *et al.*, 2024 24

# Evidence of decomposition in the 3 $ZnCl_2$ + $Li_6WN_4$ reaction

Figure S4 shows that a trace Zn impurity can be detected in reactions between $Li_6WN_4$ + 3.1 $ZnCl_2$, even when heated at low temperatures (ca. 250 °C). This impurity makes the powder appear grey in color. The Zn is likely produced via decomposition of $Zn_3WN_4$ during the highly exothermic reaction (See DSC measurements, Figure 5). Surprisingly, we do not observe W or WN. This may indicate that tungsten remains in the $Zn_3WN_4$ phase (which would then be Zn-poor), or that the W (or WN) is amorphous.

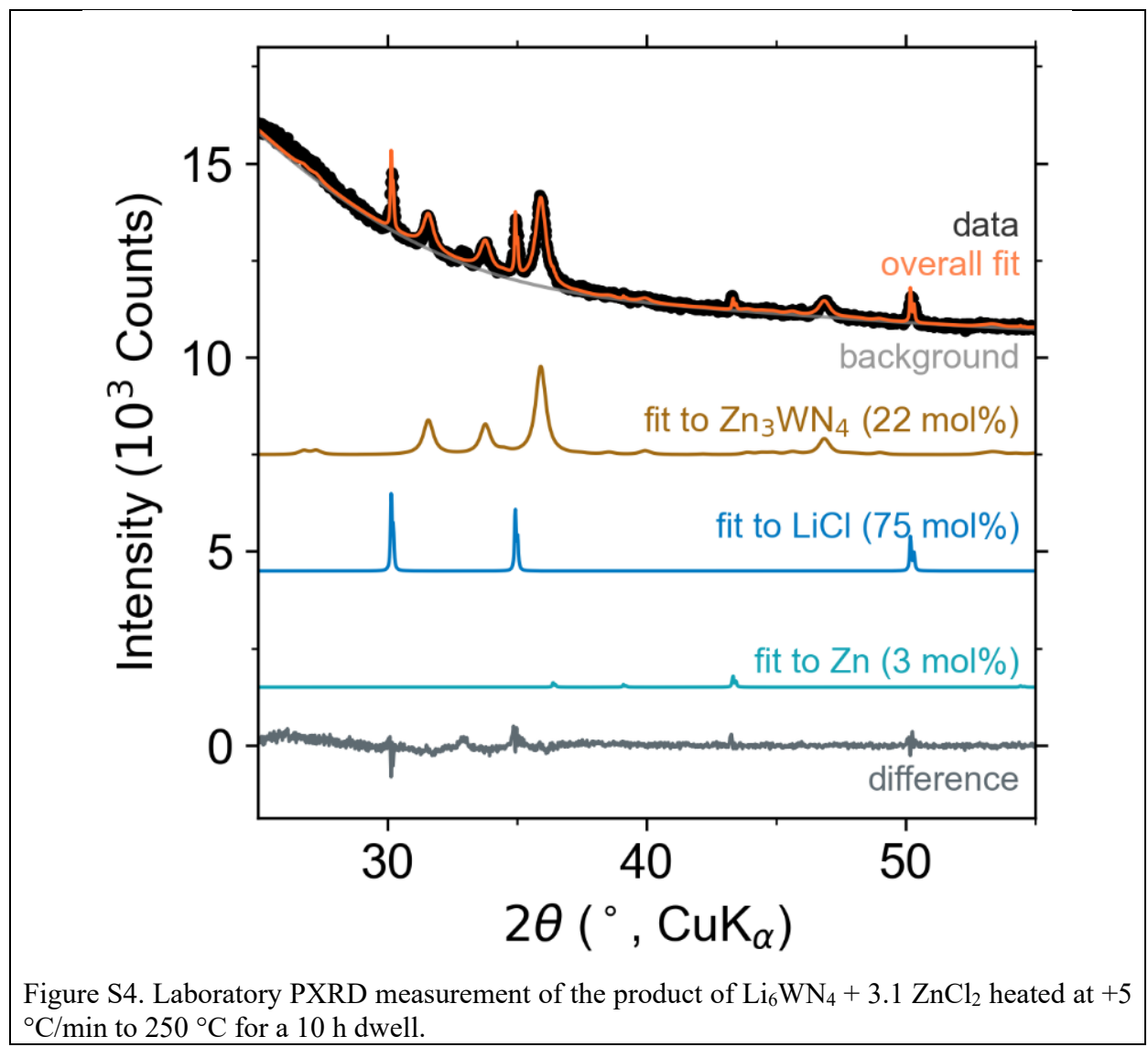

Figure S4. Laboratory PXRD measurement of the product of $Li_6WN_4$ + 3.1 $ZnCl_2$ heated at +5 °C/min to 250 °C for a 10 h dwell.

## Compositional characterization

X-ray Fluorescence (XRF) spectroscopy identified a Zn:W ratio of 3.8:1, in excess of the expected 3:1 ratio for $Zn_3WN_4$. XRF was conducted on pelletized powder after washing away the bromide byproduct. The excess zinc may be incorporated in the wurtzite-derived lattice, as suggested by Rietveld analysis (discussed below). A representative raw spectrum is shown in Figure S5. Spectra were collected at 4 different points across the pellet, and each spectrum was fit using the Bruker software to quantify Zn and W atomic ratios. The Zn:W ratio of 3.8:1 was calculated by averaging the Zn:W values from the 4 spectra.

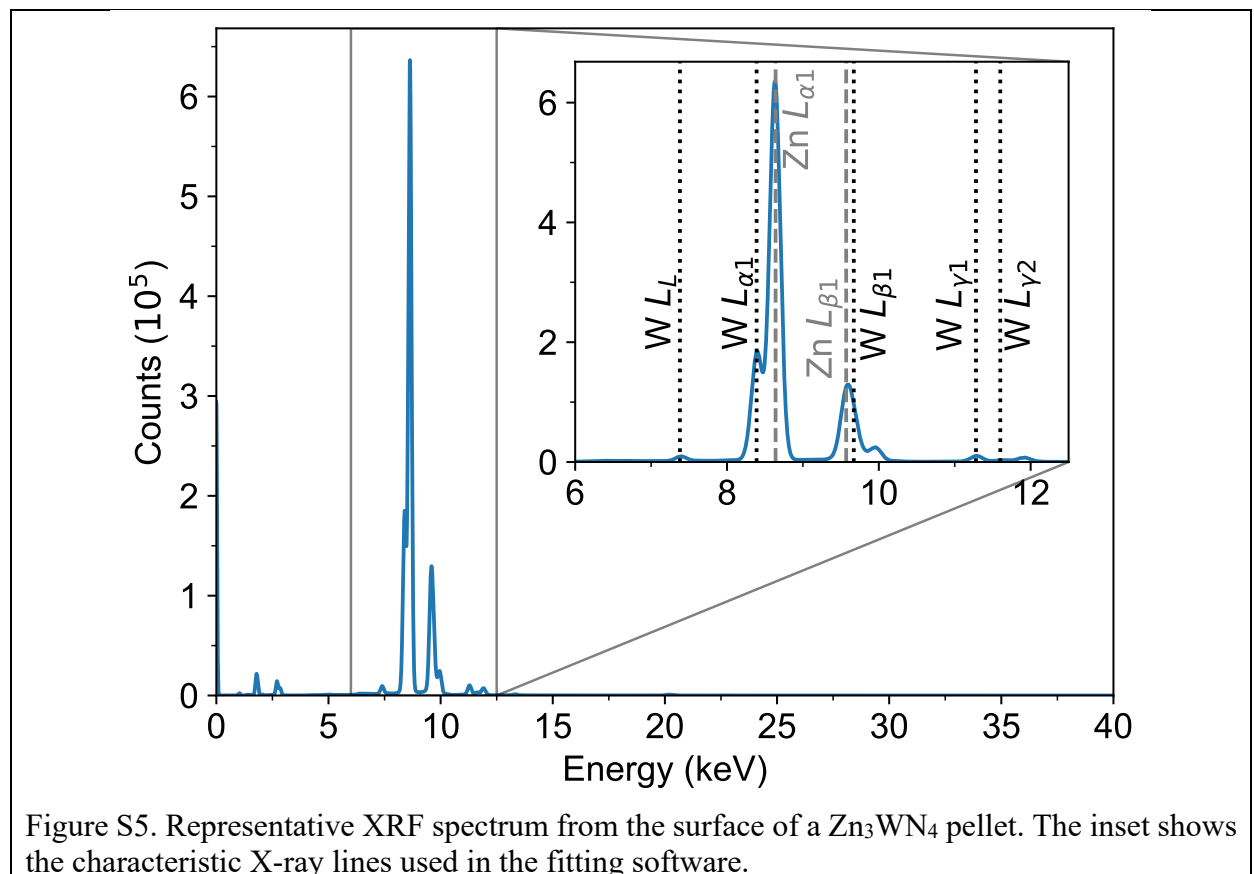

Figure S5. Representative XRF spectrum from the surface of a $Zn_3WN_4$ pellet. The inset shows the characteristic X-ray lines used in the fitting software.

# Structural models for the high resolution SPXRD measurements

The refined lattice parameters for $Zn_3WN_4$ are shown in Tables S1.

Table S1. Refined atomic coordinates for $Zn_3WN_4$ in space group $Pmn2_1$ from the SPXRD shown in Figure 1a. The unit cell parameters refined to $a$ = 6.5602(8) Å, $b$ = 5.6813(7) Å, and $c$ = 5.3235(2) Å. $R_{exp}$= 1.182 %, $R_{wp}$ = 3.989 %, $R_p$ = 2.802 %, GOF = 3.374. This structure is shown in Figure 4d.

| Site | Wyckoff | Atom | $x$ | $y$ | $z$ | Occupancy | $B_{iso}$ (Å$^2$) |
|---|---|---|---|---|---|---|---|
| Zn1 | 4b | Zn | 0.251(1) | 0.834(1) | 0.697(1) | 1.00(1) | 0.10(4) |
| Zn1 | 4b | W  | 0.251(1) | 0.834(1) | 0.697(1) | 0.00(1) | 0.10(4) |
| Zn2 | 2a | Zn | 0 | 0.337(6) | 0.698(1) | 1.00(1) | 0.20(7) |
| Zn2 | 2a | W  | 0 | 0.337(6) | 0.698(1) | 0.00(1) | 0.20(7) |
| W1  | 2a | W  | 0 | 0.670(2) | 0.221(1) | 0.83(2) | 0.44(4) |
| W1  | 2a | Zn | 0 | 0.670(2) | 0.221(1) | 0.17(2) | 0.44(4) |
| N1  | 4b | N  | 0.765(2) | 0.846(3) | 0.095(1) | 1 | 0.30(18) |
| N2  | 2a | N  | 0 | 0.330(4) | 0.116(2) | 1 | 0.4(3) |
| N3  | 2a | N  | 0 | 0.678(4) | 0.555(2) | 1 | 0.4(3) |

We considered several structural models of $Zn_3WN_4$ when conducting Rietveld analysis against our high resolution SPXRD patterns (Table S2). Several terms were allowed to vary for each approach: sample displacement, lattice parameters, size broadening (Lorentzian), strain broadening (Lorentzian), isotropic displacement parameters, and a 15-term background Chebyshev polynomial. Our most robust model was a single-phase model that allowed for a small degree of cation disorder for $Pmn2_1$ $Zn_3WN_4$, but with each cation site fixed to full occupancy (e.g., the Zn1 site was refined with Zn occupancy set to 1-$x$ and W occupancy $x$). This model resulted in an $R_{wp}$ of 3.989 % and is shown in Figure 4, Table S1, and Figure S6a. For comparison, a simpler model of $Pmn2_1$ $Zn_3WN_4$ with fixed cation occupancies (e.g., the Zn1 site fixed with 1.0 Zn occupancy) gave a significantly worse fit to the data ($R_{wp}$ = 4.638 %). Atomic positions were allowed to refine for both these single phase models. However, our diffuse reflectance spectroscopy measurements suggest that the material is not a single homogeneous phase.

Given the two distinct absorption onsets shown in the diffuse reflectance spectrum (Figure 5), we also considered a two-phase model in our Rietveld refinements. For the first two-phase model, we started with the $Pmn2_1$ $Zn_3WN_4$ from the fixed cation occupancy models. We then fixed atomic positions. Next, we created a model for cation disordered $Zn_3WN_4$ in a $P6_3mc$ structure (i.e., the wurtzite structure type), and set the lattice parameters to $a$ = 3.280 Å and $c$ = 5.324 Å such that the (100), (002), and (101) reflections of the $P6_3mc$ structure matched the (210), (002), and (211) reflections, respectively, of the $Pmn2_1$ structure (i.e., $a_{dis}$ = 0.5$a_{ord}$ and $c_{dis}$ = $c_{ord}$; where "dis" and "ord" indicate the disordered $P6_3mc$ and the ordered $Pmn2_1$ structures, respectively). This structure is consistent with the cation-disordered $Zn_3WN_4$ synthesized via thin film sputtering.[13] We then refined the size and strain broadening for both phases. This refinement resulted in 78 mol% phase fraction of $Pmn2_1$, 22 mol% for $P6_3mc$, and an $R_{wp}$ value of 3.953 % (Figure S7a), comparable to the single-phase model.

The best fit was obtained via a two-phase model, but with non-physical lattice parameters. Allowing the $P6_3mc$ lattice parameters to freely refine results in the best fit we obtained by the Rietveld method ($R_{wp}$ = 3.742 %). However, the model is likely non-physical. In this model, the refined $c$ lattice parameter for this $P6_3mc$ phase increases substantially (5.451(2) Å) compared to the $Pmn2_1$ phase ($c$ = 5.3228(2) Å), which is not consistent with prior studies of order-disorder transitions in wurtzite derived structures (e.g., $ZnGeN_2$).[14] The $c$ lattice parameters of $ZnGeN_2$ are identical in the $Pna2_1$ (the ordered structure) and $P6_3mc$ (the disordered structure), because cation disorder does not affect the layer spacing of the hcp anions along the (00$l$) direction. This analysis reveals ambiguities in these two-phase models.

Given the limitations of the two-phase models, we posit that the single phase model provides the most reliable fit to the SPXRD data without over-fitting the pattern.[15] Yet, the SPXRD measurements probe the long range average ordered structure. Local ordering—which we do not probe here—may influence the optical absorption properties shown in Figure 5. The impact of local ordering on optical properties has been characterized in the halide perovskite $CsSnBr_3$,[16] in Fe doped $SrTiO_3$,[17] and in carbon coated $FeF_3$.[18]

Table S2. Summary of models considered for the high resolution SPXRD data for $Zn_3WN_4$.

| Rietveld approach | Space group | Composition | $R_{wp}$ (%) | Figure |
|---|---|---|---|---|
| Single phase. Refined cation occ. | $Pmn2_1$ | $Zn_{3.17}W_{0.83}N_4$ | 3.989 | 1a and S6a |
| Single phase. Fixed cation occ. | $Pmn2_1$ | $Zn_3WN_4$ | 4.638 | S6b |
| Two phases. Fixed cation occ. Fixed $a$ and $c$ for $P6_3mc$ | $Pmn2_1$ $P6_3mc$ | $Zn_3WN_4$ $Zn_{0.75}W_{0.25}N$ | 3.953 | S7a |
| Two phases. Fixed cation occ. Refined $a$ and $c$ for $P6_3mc$ | $Pmn2_1$ $P6_3mc$ | $Zn_3WN_4$ $Zn_{0.75}W_{0.25}N$ | 3.742 | S7b |

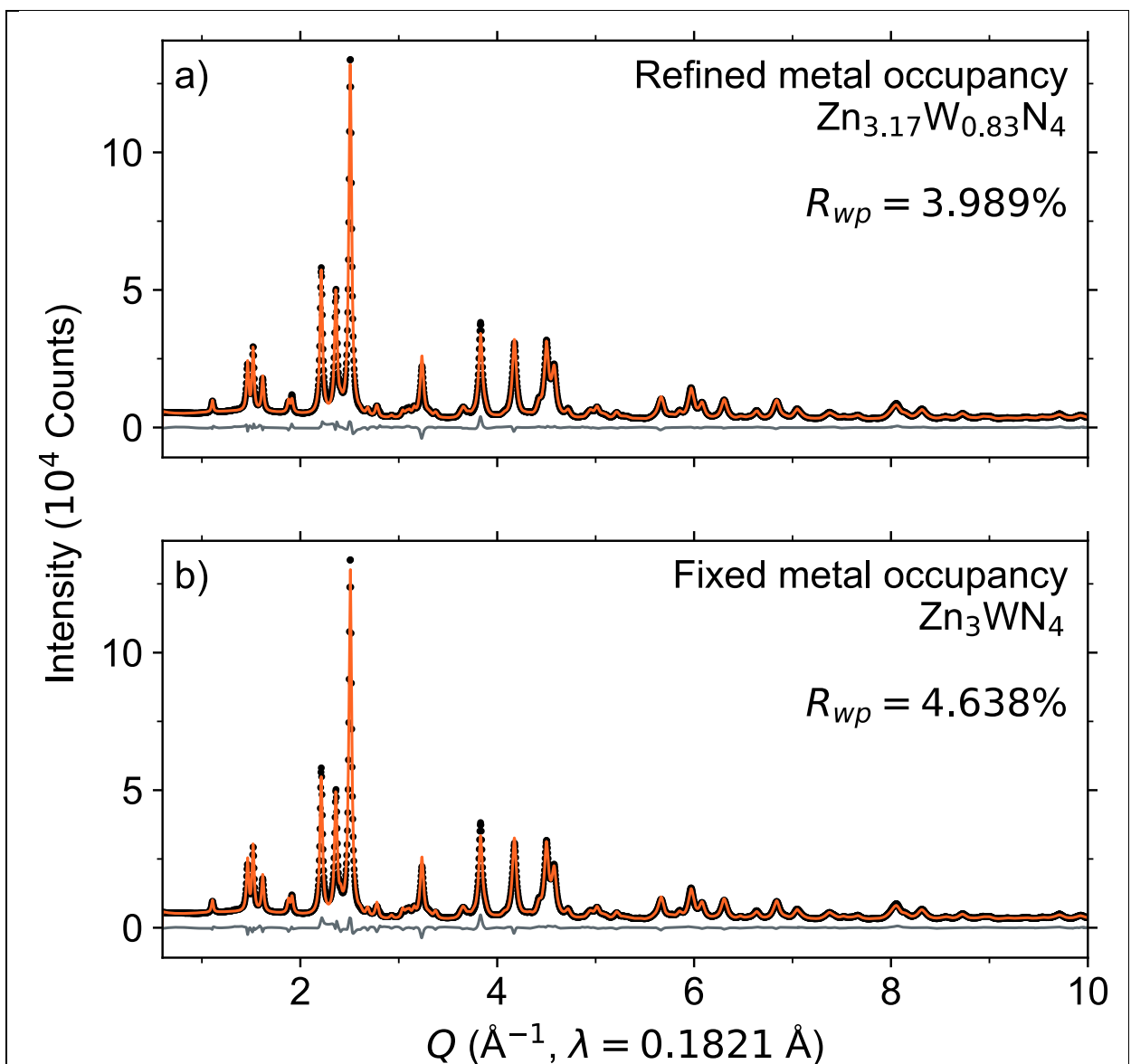

Figure S6. The high resolution SPXRD measurement of the $Zn_3WN_4$ sample fit with single-component models of $Pmn2_1$ with a) the occupancies of the metal sites refined to $Zn_{3.17}W_{0.83}N_4$ (also shown in Figure 4), and b) fixed metal site occupancy at cation-ordered $Zn_3WN_4$.

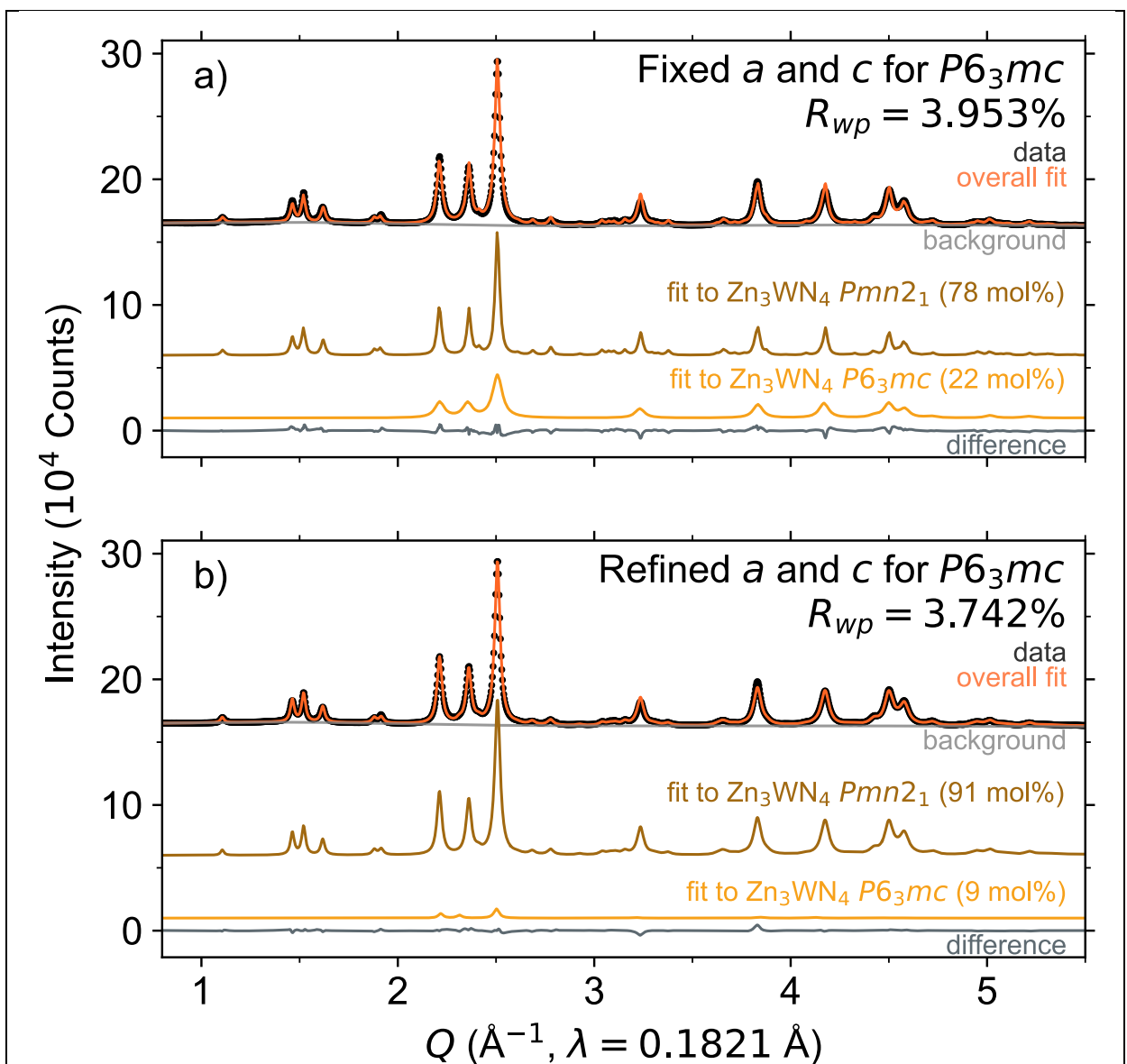

Figure S7. Rietveld refinement of the high-resolution SPXRD data of $Zn_3WN_4$ using a two-component model with a) the $P6_3mc$ lattice parameters fixed relative to the $Pmn2_1$ values and b) the $P6_3mc$ lattice parameters freely refined.

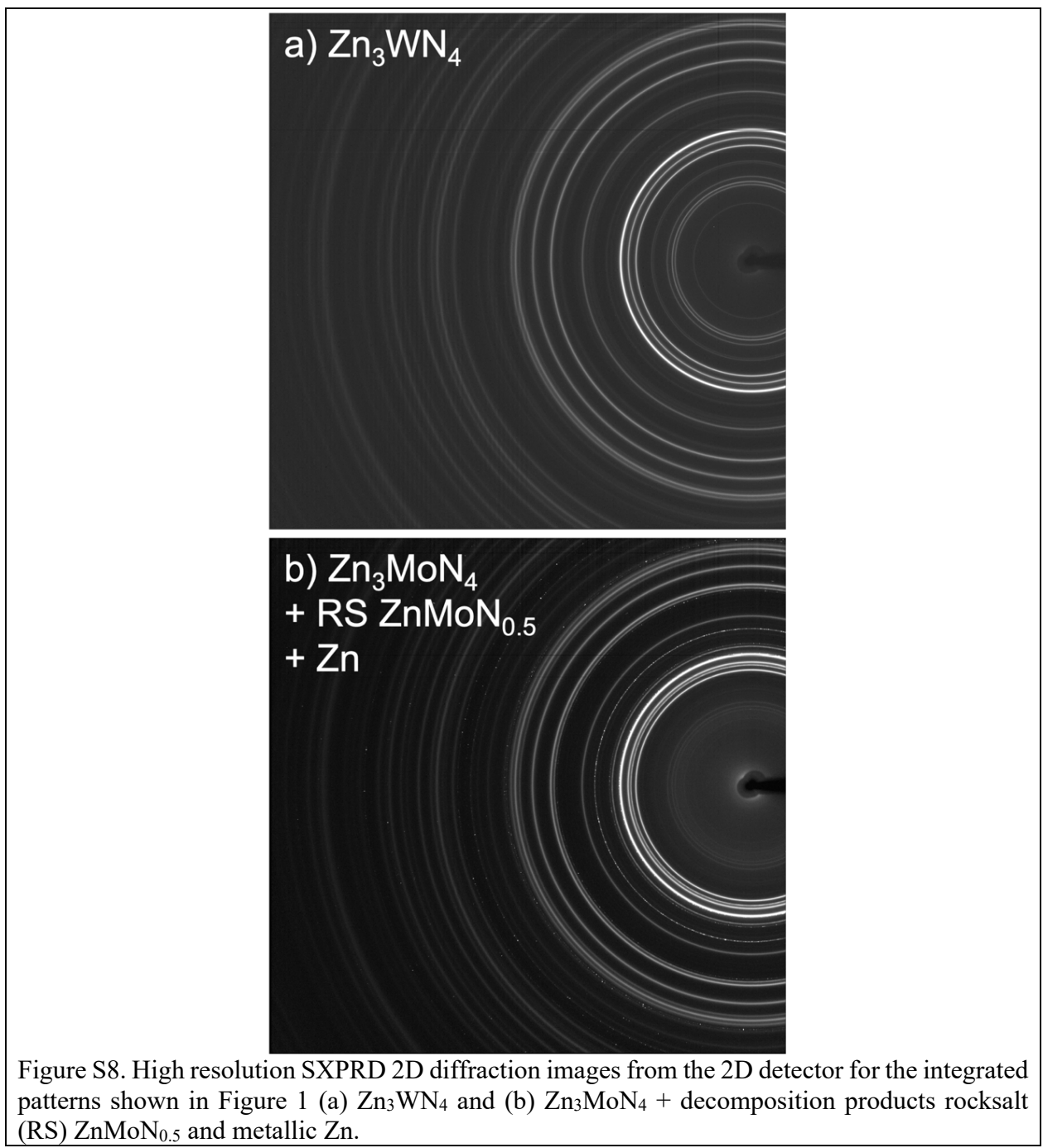

Figure S8. High resolution SXPRD 2D diffraction images from the 2D detector for the integrated patterns shown in Figure 1 (a) $Zn_3WN_4$ and (b) $Zn_3MoN_4$ + decomposition products rocksalt (RS) $ZnMoN_{0.5}$ and metallic Zn.

Qualitative inspection of the 2D diffraction images show that the diffraction rings for the $Zn_3WN_4$ powder are homogeneous (Figure S8). If rings with two different morphologies were present, this would suggest the presence of two distinct phases with different crystallinity, size, and strain. That we see only one morphology of ring in the 2D detector image supports either a single phase or multiple phases with nearly identical crystallinity, size, and strain. In our Rietveld refinement that modeled the data using two phases, the fit to the data is significantly worse when we constrain the size and strain broadening terms to be the same for both phases. These findings support our use of the single-phase model.

## Magnetic susceptibility measurements

Magnetic measurements were performed on $Zn_3WN_4$ using a Quantum Design Physical Property Measurement System (PPMS). A powder sample of $Zn_3WN_4$ was loaded into a small packet (0.001125 mg) and secured inside a plastic straw for the measurement. Magnetic susceptibility ($\chi$) of $Zn_3WN_4$ was measured as a function of temperature shows largely diamagnetic behavior with a trace paramagnetic impurity (Figure S9a). Similarly, magnetization ($M$) as a function of applied field ($H$) at 2 K shows that diamagnetism dominates the field-dependent magnetization (Figure S9b). These findings are inconsistent with pure $Zn_3WN_4$, which should be purely diamagnetic. Zn impurities, if present, would also give a diamagnetic response. The paramagnetic component suggests the possibility of a reduced tungsten species (e.g., $W^{5+}$), possibly as a sub-nitride (e.g., $Zn_3WN_{4-\delta}$), an oxynitride impurity (e.g., $Zn_3WN_{4-x}O_x$), or a W-rich phase (e.g., $Zn_{3-\delta}W_{1+\delta}N_4$).

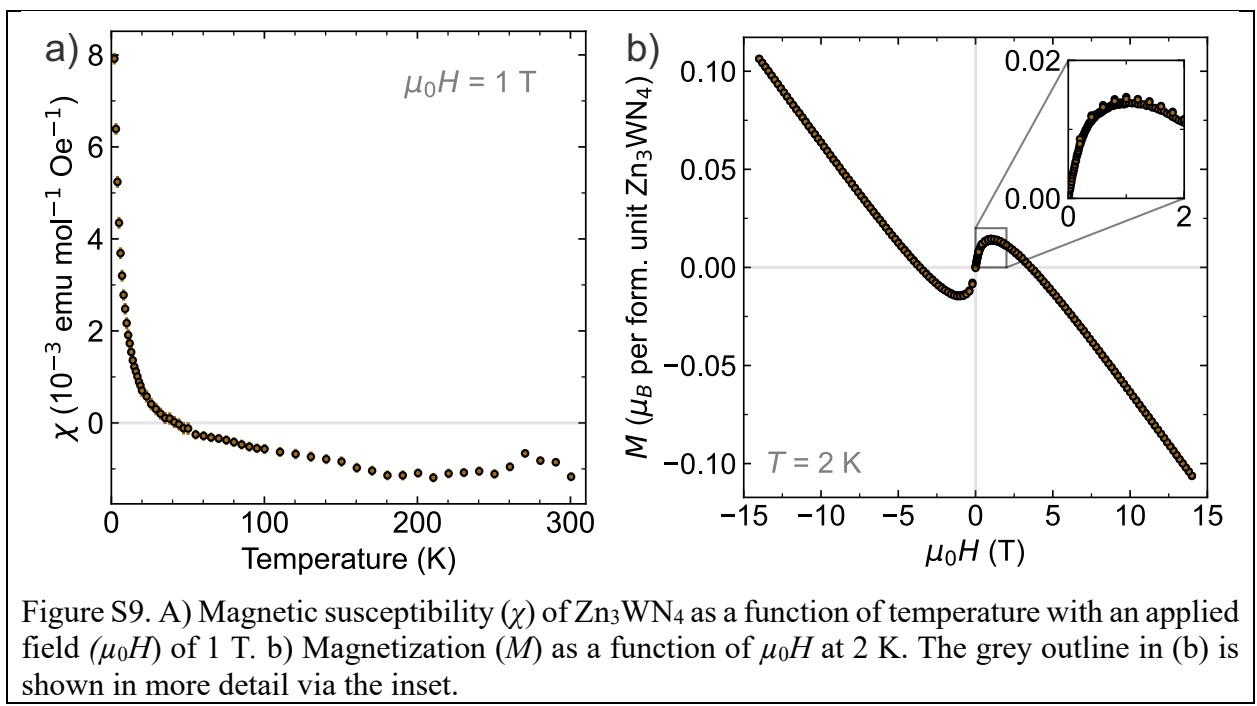

Figure S9. A) Magnetic susceptibility ($\chi$) of $Zn_3WN_4$ as a function of temperature with an applied field ($\mu_0H$) of 1 T. b) Magnetization ($M$) as a function of $\mu_0H$ at 2 K. The grey outline in (b) is shown in more detail via the inset.

# Full structure visualizations of Li$_6$WN$_4$ and Zn$_3$WN$_4$

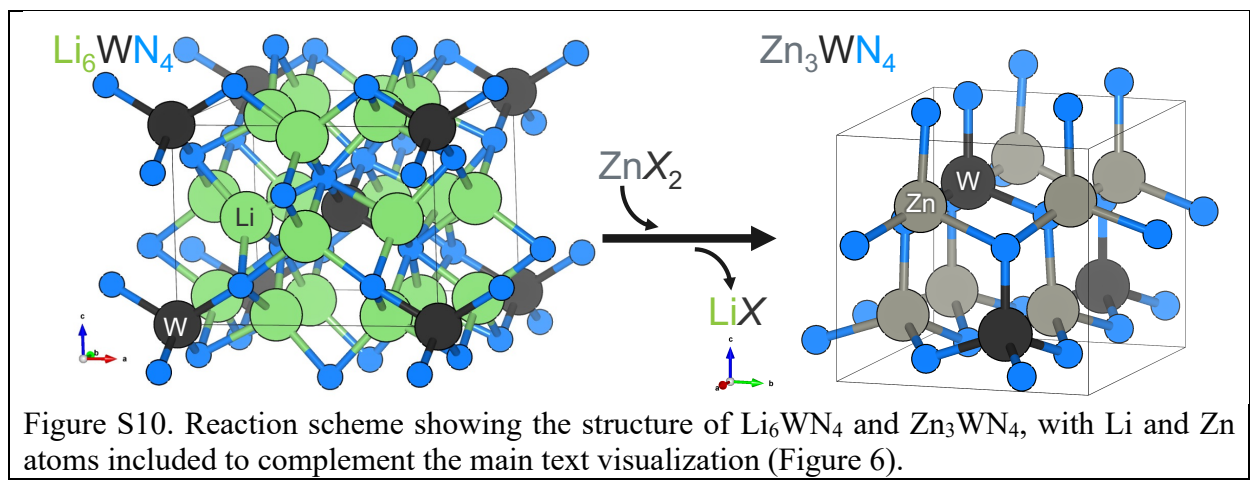

Figure S10. Reaction scheme showing the structure of Li$_6$WN$_4$ and Zn$_3$WN$_4$, with Li and Zn atoms included to complement the main text visualization (Figure 6).

# Synthesis of Zn$_3$MoN$_4$

The synthesis strategy used for Zn$_3$WN$_4$ was also applied to synthesize Zn$_3$MoN$_4$, but the product exhibited partial decomposition. Analysis of the SPXRD pattern collected for the Mo analog of Li$_6$WN$_4$, synthesized via the reaction Li$_6$MoN$_4$ + 3 ZnBr$_2$ → Zn$_3$MoN$_4$ + 6 LiBr, suggests phase decomposition (Figure S11). In addition to the desired Zn$_3$MoN$_4$ (56 mol%), Rietveld analysis of High resolution SPXRD data show that a rocksalt (RS) structure fit as ZnMoN$_{0.5}$ also forms (23 mol%), along with a Zn impurity (20 mol%). This RS phase exhibits a substantially larger lattice parameter ($a$ = 4.7106(3) Å) than the defect-RS phase Mo$_2$N (a = 4.16 Å to 4.19 Å).[19,20] Therefore, we hypothesize it may be a (Zn,Mo)N$_x$ material, as octahedra Zn$^{2+}$ has a substantially larger ionic radius (0.74 Å) than octahedral Mo$^{x+}$ (0.65 Å for Mo$^{4+}$, 0.69 Å for Mo$^{3+}$).[21] The rocksalt ZnMoN$_{0.5}$ phase was created from a $Fm\bar{3}m$ Mo$_2$N starting model. Rietveld analysis with the composition of ZnMoN$_{0.5}$ provides a reasonable fit. Further analysis of this material is beyond the scope of this manuscript. Additional minor peaks that we have not indexed are present (possibly higher order oxides). These impurity phases suggest that Zn$_3$MoN$_4$ is less stable at elevated temperatures than Zn$_3$WN$_4$. This decomposition occurs despite the excess ZnBr$_2$ which was intended to serve as a heat sink during the exothermic reaction. Despite the partial decomposition of the phase, Zn$_3$MoN$_4$ is still the major phase in the pattern. As with Zn$_3$WN$_4$, the SPXRD pattern for Zn$_3$MoN$_4$ shows evidence of cation-ordering in the form of the $Pmn2_1$ reflections at low angle: e.g., (010), (110), (101), (011). However, these reflections are weaker than in the W case, owing to the lower scattering factor of Mo compared to W. We focused our work on Zn$_3$WN$_4$ because W scatters X-rays more strongly than Mo (facilitating characterization) and because our Zn$_3$WN$_4$ products exhibited higher phase purity.

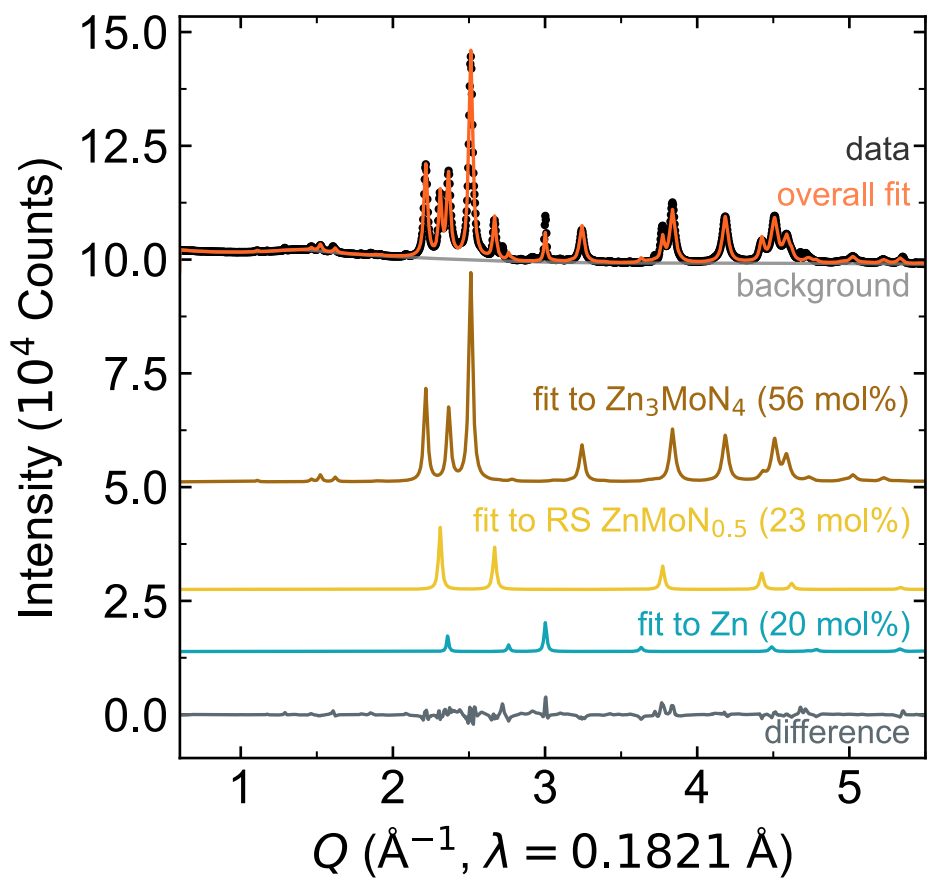

Figure S11. SPXRD pattern and Rietveld refinement of the washed products from the reaction between $Li_6MoN_4$ + 4.2 $ZnBr_2$ (excess $ZnBr_2$).